\documentclass[electronic]{vgtc}             % electronic version

\ifpdf%                                % if we use pdflatex
  \pdfoutput=1\relax                   % create PDFs from pdfLaTeX
  \usepackage{graphicx}                % allow us to embed graphics files
  \DeclareGraphicsExtensions{.pdf,.png,.jpg,.jpeg} % for pdflatex we expect .pdf, .png, or .jpg files
\else%                                 % else we use pure latex
  \usepackage{graphicx}                % allow us to embed graphics files
  \DeclareGraphicsExtensions{.eps}     % for pure latex we expect eps files
\fi%

\graphicspath{{figures/}{pictures/}{images/}{./}} % where to search for the images

\usepackage{microtype}                 % use micro-typography (slightly more compact, better to read)
\PassOptionsToPackage{warn}{textcomp}  % to address font issues with \textrightarrow
\usepackage{textcomp}                  % use better special symbols
\usepackage{mathptmx}                  % use matching math font
\usepackage{times}                     % we use Times as the main font
\usepackage{cite}                      % needed to automatically sort the references
\usepackage{tabu}                      % only used for the table example
\usepackage{booktabs}                  % only used for the table example
\usepackage{amsfonts,amsthm}
\usepackage{array}
\usepackage{booktabs}
\usepackage{color}
\usepackage{enumitem}
\usepackage{float}
\usepackage{mathtools}
\usepackage{multirow}
\usepackage{setspace}
\usepackage{url}

\theoremstyle{definition}

\newcolumntype{L}[1]{>{\raggedright\let\newline\\\arraybackslash\hspace{0pt}}m{#1}}
\newcolumntype{C}[1]{>{\centering\let\newline\\\arraybackslash\hspace{0pt}}m{#1}}
\newcolumntype{R}[1]{>{\raggedleft\let\newline\\\arraybackslash\hspace{0pt}}m{#1}}
\newcommand\Tstrut{\rule{0pt}{2.6ex}}         % = `top' strut
\setlist[itemize]{noitemsep, topsep=0pt}
\setlist[enumerate]{noitemsep, topsep=0pt}

\newcommand{\parens}[1]{\left(#1\right)}
\newcommand{\braces}[1]{\left\{#1\right\}}
\newcommand{\bracks}[1]{\left[#1\right]}

\newcommand{\norm}[1]{\left\Vert#1\right\Vert}

\newcommand{\raisedth}[1]{#1^{\textrm{th}}}
\onlineid{1209}

\vgtccategory{Research}

\vgtcinsertpkg

\title{Generating Emotive Gaits for Virtual Agents Using Affect-Based Autoregression}

\author{Uttaran Bhattacharya\thanks{e-mail: \url{uttaranb@umd.edu}}\\ %
        \scriptsize UMD College Park, USA %
\and Nicholas Rewkowski\thanks{e-mail: \url{nick1@umd.edu}}\\ %
        \scriptsize UMD College Park, USA, UNC Chapel Hill, USA %
\and Pooja Guhan\thanks{e-mail: \url{pguhan@umd.edu}}\\ %
        \scriptsize UMD College Park, USA %
\and Niall L. Williams\thanks{e-mail: \url{niallw@umd.edu}}\\ %
        \scriptsize UMD College Park, USA %
\and Trisha Mittal\thanks{e-mail: \url{trisha@umd.edu}}\\ %
        \scriptsize UMD College Park, USA %
\and Aniket Bera\thanks{e-mail: \url{ab@cs.umd.edu}}\\ %
        \scriptsize UMD College Park, USA %
\and Dinesh Manocha\thanks{e-mail: \url{dm@cs.umd.edu}}\\ %
        \scriptsize UMD College Park, USA}

\teaser{
    \centering
    \includegraphics[width=0.7\textwidth]{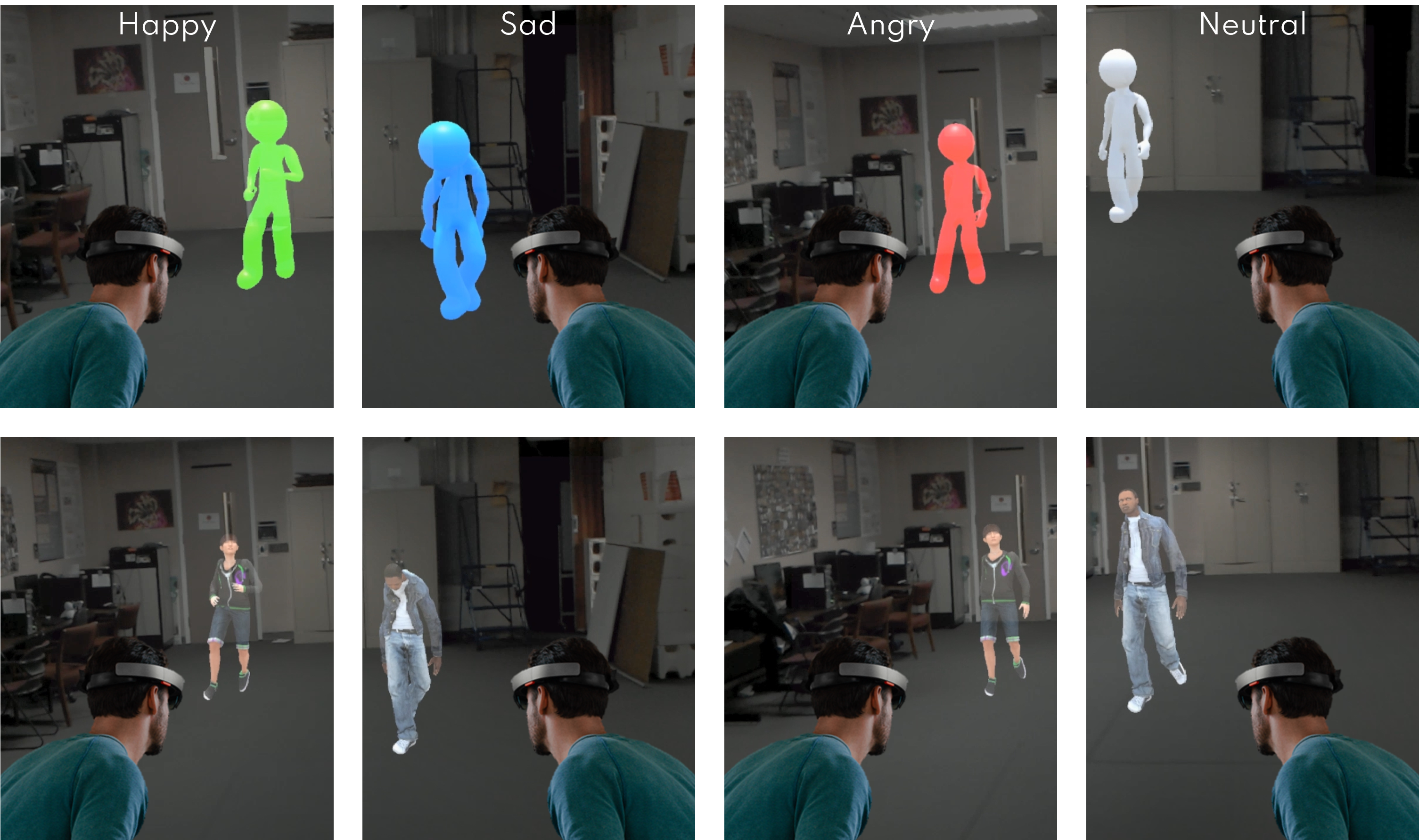}
    \caption{\textbf{Emotive Gaits for Virtual Agents (VAs) Generated Using Affect-Based Autoregression:} Samples of the emotive gaits for VAs generated by our learning-based algorithm in an AR environment. The top row shows stick figures and the bottom row shows human models corresponding to the VAs. The VAs can exhibit different emotions  based on their gaits, as indicated at the top of each column, while walking along user-defined trajectories at interactive rates.}
    \label{fig:teaser}
}

\abstract{
We present a novel autoregression network to generate virtual agents that convey various emotions through their walking styles or gaits. Given the 3D pose sequences of a gait, our network extracts pertinent movement features and affective features from the gait. We use these features to synthesize subsequent gaits such that the virtual agents can express and transition between emotions represented as combinations of happy, sad, angry, and neutral. We incorporate multiple regularizations in the training of our network to simultaneously enforce plausible movements and noticeable emotions on the virtual agents. We also integrate our approach with an AR environment using a Microsoft HoloLens and can generate emotive gaits at interactive rates to increase the social presence. We evaluate how human observers perceive both the naturalness and the emotions from the generated gaits of the virtual agents in a web-based study. Our results indicate around 89\% of the users found the naturalness of the gaits satisfactory on a five-point Likert scale, and the emotions they perceived from the virtual agents are statistically similar to the intended emotions of the virtual agents. We also use our network to augment existing gait datasets with emotive gaits and will release this augmented dataset for future research in emotion prediction and emotive gait synthesis. Our project website is available at {\color{red}\url{https://gamma.umd.edu/gen_emotive_gaits/}}.}
\CCScatlist{
  \CCScatTwelve{Human-centered computing}{Human computer interaction (HCI)}{Interaction paradigms}{Mixed / augmented reality};
  \CCScatTwelve{Computing methodologies}{Machine learning}{Machine learning approaches}{Neural networks}
}

\begin{document}


\firstsection{Introduction}\label{sec:intro}

\maketitle

%% \section{Introduction} %for journal use above \firstsection{..} instead

%% Intro
The generation of intelligent virtual agents (IVAs) is important for many virtual and augmented reality systems. The virtual agents correspond to embodied digital characters that are often used as avatars to represent the users and may look like real-world characters. Recent work in photorealistic rendering and capturing technologies has resulted in generating agents or avatars that closely resemble the humans and are widely used in VR and AR systems~\cite{neon,rocketbox}.

Many applications, including virtual assistance, training, and AI chatbots, need a computationally created virtual agent or an avatar that not only looks like a real human but also behaves like one and conveys emotions~\cite{avatar_realism,fva}.  Perception of such emotional expressiveness is commonly described as the ability of an observer to make decisions on a subject's emotional state by observing certain patterns or cues physically expressed by the subject~\cite{neurobiology_of_perception,critical_gait_features,effort_shape}. These physical cues are expressed through various ``modalities,'' including, but not limited to, facial expressions~\cite{face_ex}, tones of voice~\cite{voice_ex}, gestures, body expressions~\cite{posture_ex}, walking styles or ``gaits''~\cite{montepare}, etc. Emotions coming from such different modalities~\cite{m3er}, in conjunction with the different underlying situational and social contexts~\cite{context_based_ER,emoticon}, have a
significant impact on our everyday lives. They influence our social interactions and relationships and provide key insights into developing healthy social environments~\cite{emotions_social}. Similarly, 
emotions can greatly impact the perception of these virtual agents in terms of social presence and how the users behave when interacting with them in AR and VR environments ~\cite{embodied_ambient_crowds,small_group_interaction,fva}.  

In this paper, we mainly focus on designing virtual agents that are capable of expressing different emotions through their gaits, \textit{i.e.} virtual agents with {\em emotive gaits}.
When perceiving emotions from gaits, humans generally look at physical expressions such as arm swing, stride length, upper-body posture, head jerk, etc.~\cite{crenn2016body}, collectively referred to as \textit{affective features}. In fact, studies have shown that observers often rely on such cues from gaits and other body expressions, especially when there are mismatches with cues from more common modalities such as facial expressions~\cite{misleading_cues}. As a result, it is useful to generate virtual agents with emotive gaits for gaming and social VR~\cite{games,vr}, crowd simulation, and path planning~\cite{robotics,proxemo}, anomaly detection~\cite{liarswalk}, therapy, rehabilitation~\cite{therapy}, and psychology and neurobiology~\cite{depression,context_PS,brain_injury}. Studies indicate that virtual agents expressing various moods, emotions, and behaviors elicit more empathy and engagement from humans interacting with them~\cite{anthropomorphism}. However, automated methods to generate gaits that express certain emotions are challenging to design and implement. This is hard not only because of the complexity of modeling periodic and aperiodic motions constituting different gaits, but also because of the individual, social, and cultural diversities in terms of both expressing and perceiving emotions~\cite{emotion_culture_diversity1,emotion_culture_diversity2}. These challenges are further exacerbated by the difficulty of collecting and annotating large benchmark datasets of gaits with appropriate emotion labels. 
Datasets collected in controlled experimental settings are often exaggerated and not always generalizable. On the other hand, datasets collected in the wild often suffer from the long-tail nature of the distribution of emotions, with most emotion being close to neutral. Therefore, there is a need to develop automated methods to synthesize emotive gaits for virtual agents that can augment the existing datasets.

There is an extensive work in AR/VR, computer graphics, vision and related areas such as biomechanics, on automated techniques for generating human characters capable of performing locomotion activities, including walking, running, leaping, and more~\cite{dominance,eva,fva,motion_synth,pfnn,deep_loco,spectral_style_transfer}.
These methods make use of movement features such as joint rotations, joint velocities, frequency of foot contact with the ground, and walking phases, and combine them with structural and kinematic constraints of the human body and learning techniques. While these methods can generate plausible locomotion, walking patterns, and actions, it is non-trivial to add emotional components to these techniques or use them to generate emotive gaits.

\noindent {\bf Main Results:}
We present a novel autoregression method that takes as input 3D pose sequences of the gaits of virtual agents (VAs) and efficiently combines pose affective features such as arm swings, head jerks, body posture and more, and movement features such as stepping speed, root height and more, to generate future pose sequences of emotive gaits. We present a network architecture that incorporates both spatial information and spectral information available from the input pose sequences and enables the VAs to both express and transition smoothly between different emotions while walking. We construct VAs as both stick figures and human models using our generated emotive gaits and integrate these VAs in an AR environment using the Microsoft HoloLens (Figure~\ref{fig:teaser}). Our learning-based algorithm takes a few milliseconds to generate an emotive gait for each agent on a pair of NVIDIA GeForce GTX 1080Ti GPUs. Our VAs, overlayed onto a real-world room, are rendered at interactive rates to
increase their sense of social presence.
The novel components of our work include:
\begin{itemize}
    \item An autoregression network that takes in 3D pose sequences of a VA's gait, the desired future trajectory, and the desired emotions. It outputs the VA's gait expressing the given emotion while following the given trajectory.
    \item A novel training method combining movement features and psychologically-motivated affective features into a unified network to generate plausible, emotionally-expressive gaits.
    \item A transition scheme for the characters to smoothly transition between gaits expressing different emotions.
    \item An elaborate web-based user study to evaluate the benefits of the emotive gaits generated by our algorithm. We asked the observers to report the emotions they perceived from the generated gaits, as well as the Likert scale (LS) values of pose affective features that contributed to their perception. Based on the study, we conclude that
    \begin{itemize}
        \item  There is strong statistical evidence to suggest that the observers' perceived emotions are statistically similar to the corresponding intended emotions of our VAs, thereby showing that the generated gaits are emotive,
        \item The observers consistently reported different LS values of the pose affective features for different emotions, making our choice of pose affective features statistically significant for perceiving emotional expressiveness.
    \end{itemize}
    \item An augmented dataset, ``Synthesized Emotive Gait,'' which provides emotive gaits generated by our method to facilitate more research in this area.
\end{itemize}

%%Related Work
\section{Related Work}\label{sec:rw}
In this section, we briefly survey prior work on representing emotions, perceiving emotions from gaits, generating and styling gaits for virtual agents, and making virtual agents emotionally expressive.

\subsection{Modeling and Perceiving Emotions from Gaits}\label{subsec:emotions_from_gait}
Various models for representing emotions have been studied in psychology, and the Valence-Arousal-Dominance (VAD) model~\cite{vad} is one of the most popular. The VAD model considers a continuous 3D space, spanned by the valence, arousal, and dominance axes. Valence is a measure of the pleasantness of emotion, arousal is a measure of the intensity of expression, and dominance is the measure of how much emotion makes one feel in control. Many methods use a simpler model that is a linear combination of discrete emotions and represents a subset of VAD.

Humans perceive these emotions by observing physical features or cues expressed via different modalities. 
Studies conducted by Montepare et al.~\cite{montepare} concluded that observers were able to perceive emotions by only looking at the subjects' gaits. Subsequently, Roether et al.~\cite{critical_gait_features} and Gross et al.~\cite{effort_shape} identified that observers were most consistent when looking at gaits expressing emotions that varied on the arousal axis. 
Follow-up studies looked more closely at the gait-based expressions observers focused on for distinguishing between different perceived emotions and identified features including arm swing, gait velocity, upper body posture, and head jerk~\cite{karg2013body,crenn2016body,step,taew}.
In contrast to prior approaches, our goal is to use affective features and movement features to synthesize gaits with emotions varying on the arousal axis.

\subsection{Emotional Expressiveness in Virtual Agents}
Prior works have commonly explored the generation of emotionally expressive virtual agents via modalities such as verbal communication~\cite{verbal_comm1,verbal_comm2}, face movements~\cite{expressive_face}, body gestures~\cite{bonding_in_conversations}, and gaits~\cite{dominance,fva,eva}. These generation techniques have had significant performance benefits when combined with concepts from affective computing. For example, Pelczer et al.~\cite{evaluate_recognition} designed a strategy to evaluate the accuracy of identifying the modalities of emotional expressiveness in a virtual agent. McHugh et al.~\cite{body_posture_crowds} explored how different body postures influenced the emotion perception of individual agents in crowds. Clavel et al.~\cite{face_and_posture} studied the combined effect of faces and postures of virtual agents on emotion perception, and Liebold et al.~\cite{generalized_combinations} generalized this to include combinations of other modalities such as verbal cues and faces. More recently, Randhavane et al.~\cite{eva} developed an empirical mapping between gait and gaze features and different emotions to generate emotionally expressive virtual agents. Our approach to generating emotive gaits is complementary to these methods and can be combined with them.

\subsection{Generating and Styling Gaits for Virtual Agents}
There has been extensive prior work in computer graphics and AR/VR for generating and styling gaits for virtual agents. Early approaches used patch-based building blocks~\cite{motion_patches}, kernel-based approaches~\cite{kernel_based2}, or modeled the motion paths as directed graphs~\cite{motion_graphs} to generate natural-looking movement styles. Recent approaches have leveraged large-scale datasets using deep learning-based approaches to generate diverse movement styles. These approaches include training a network on specific joint trajectories~\cite{biped,motion_synth}, using periodic phase-functions, which are either modeled geometrically~\cite{pfnn} or learned with a neural network~\cite{nsm}, to represent walking cycles, and exploiting transfer to reduce over-dependency on data~\cite{few_shot_homogeneous}. Other approaches use deep reinforcement learning to learn control policies for virtual characters exhibiting different movement styles and actions~\cite{muscleactuation,icc,deep_loco,deep_mimic}. Yet other approaches model motion prediction as an autoregression problem, and have utilized recurrent networks~\cite{quaternet} and convolutional networks~\cite{seq2seq_dynamics} on motion captures pose sequences, and generative adversarial learning on dynamic pose graphs~\cite{dynamic_prediction}, to predict future motions. While these methods are not built for motion styling, their key concepts have been useful in developing many motion-based style transfer methods~\cite{style_transfer_heterogeneous,style_cvae,unpaired_style_transfer}.

In contrast with these methods, we combine walking phases and gait-based affective features in an autoregression network to estimate future joint rotations and movement features for different emotive gaits. Instead of a DRL-based control policy, our network learns a feature-based latent representation space and maps from that space to emotion-styled predicted poses. Furthermore, our emotion styles are sampled from a continuous space of emotions. Therefore, they need to be modeled differently from conventional styles, which can be viewed as one-hot labels in a discrete space. We also demonstrate that our approach can generate gaits expressing a continuous range of emotions, for AR applications.

\begin{figure}[t]
    \centering
    \includegraphics[width=0.7\columnwidth]{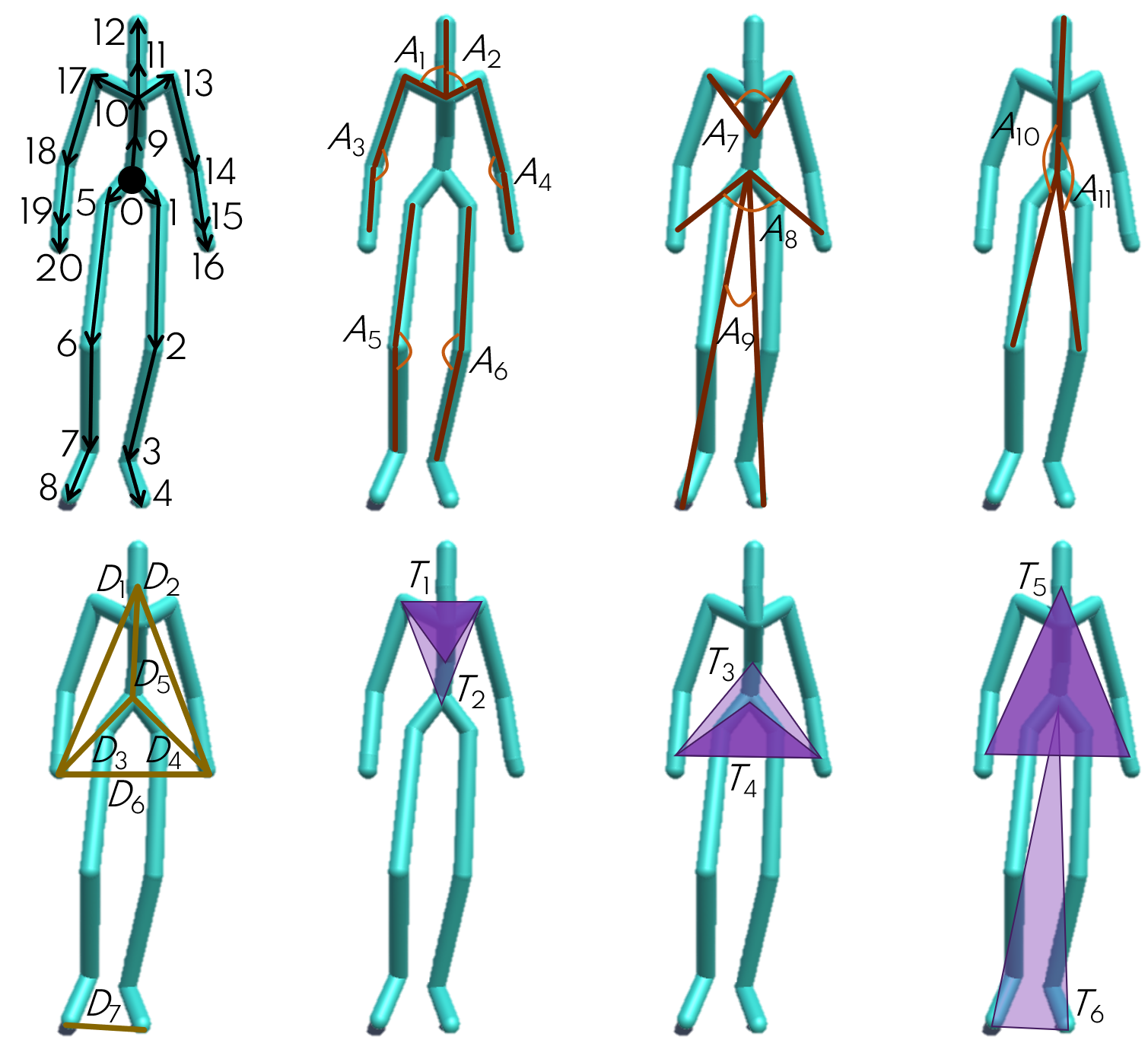}
    \caption{\textbf{Our Affective features:} The top left figure shows our pose graph as a directed tree, with the joints numbered $0$ through $20$. We use $18$ affective features, counting $11$ joint angles, $4$ distance ratios, and $3$ area ratios. The joint angles are labeled $A_1$ through $A_{11}$, and marked with red arcs on the last three figures in the top row. The leftmost figure on the bottom row shows the distances we use to compute the distance ratios. We use the ratios $\frac{D_1}{D_2}$, $\frac{D_3}{D_4}$, $\frac{D_6}{D_5}$, and  $\frac{D_7}{D_5}$. The last three figures on the bottom row show the triangles we use to compute area ratios. We use the ratios $\frac{T_1}{T_2}$, $\frac{T_3}{T_4}$, $\frac{T_5}{T_6}$. These features are used by our network to generate emotive gaits of the virtual agents.}
    \label{fig:affective_features}
    \vspace{-10pt}
\end{figure}

%% Overview
\section{Generating Emotive Gaits}\label{sec:overview}

In this section, we present our approach for generating emotive gaits of VAs. The inputs to our algorithm are the sample gaits of a VA provided as motion capture data, the desired trajectory, and the desired emotion. Our goal is to generate subsequent predictions of the gait that follow the desired trajectory and express the desired emotion. Since our approach depends only on the input motion-captured gait samples and the desired trajectories, we can adapt to VAs with different skeletal dimensions, different natural walking styles, as well as to different AR environments.

\subsection{Emotion Model}
We choose to use an emotion model consisting of linear combinations of categorical emotions varying primarily on the arousal axis (happy, angry, sad, etc.). Although this model admittedly spans a smaller set of emotions than the VAD model, prior works have reported that categorical emotion terms are more easily understood by non-experts, leading to the availability of more labeled data and the generation of a diverse range of emotions~\cite{tanmay_emotions,step}.

\subsection{Construction of the Generative Components}\label{subsec:generative}
We present various components used by our network to perform generation: input gaits, emotions, pose affective features, and trajectory features.

\subsubsection{Gaits}
We denote a gait $G$ as $G = \braces{X_j^t=\bracks{x_j^t, y_j^t, z_j^t}^\top \in \mathbb{R}^3}_{j=0, t=0}^{J-1, T-1}$, where $\bracks{x_j^t, y_j^t, z_j^t}^\top$ denotes the 3D positions of joint $j$ at time step $t$ in the world frame, $J$ denotes the total number of joints, and $T$ denotes the total number of input time steps. The input to our network are joint rotations extracted from the gait $G$, and control signals obtained from the gait, its trajectory, and the associated emotions.

At each time step $t$, we model the pose graph of the input gait as a directed tree, as shown in Figure~\ref{fig:affective_features}. The root joint in the pose is the root node of the tree and the two toes, the two hand indices, and the head, are the leaf nodes the tree. All edges in the tree are directed from the root node to the leaf nodes. We denote the parent of a joint $j$ as $P\parens{j}$. In our construction, each joint has a unique parent, except the root joint, which has no parent. We therefore assign $P\parens{j} = -1$ for the root joint. For each joint $j$ at each time step $t$, we consider the rotation $R_j^t \in \mathbb{SO}\parens{3}$ that transforms the joint from its offset $o_j \in \mathbb{R}^3$ --- a pre-defined initial position relative to its parent $P\parens{j}$ --- to its position at time step $t$ relative to its parent. That is, we consider the rotation $R_j^t$ such that the global position $X_j^t$ of the joint $j$ at time step $t$ is given by $X_j^t = R_j^t o_j + X_{P\parens{j}}^t$. For the root joint, we consider $o_0 = \mathbf{0}$ and obtain its position directly from the gait $G$.

Following the approach taken by Pavllo et al.~\cite{quaternet}, we represent these rotations as unit quaternions or versors $q_j^t \in \mathbb{H} \subset \mathbb{R}^4$, where $\mathbb{H}$ denotes the space of versors. We have chosen to represent rotations as versors, as they are free of the gimbal-lock problem.
%unlike other common representations such as Euler angles or exponential maps~\cite{exponential_maps}. 
We enforce the additional unit norm constraints for these versors when training our network.
Thus, for each gait, our input rotations are given as
$$\braces{q^t = \bracks{q_1^t; \dots; q_{J-1}^t} \in \mathbb{H}^{J-1}}_{t=0}^{T-1}.$$ We do not consider $q_0^t$ for the root joint denoted by $0$ as part of the input data, since $o_0 = 0$.

\subsubsection{Emotions}\label{subsubsec:emotions}
We note that the modeling of emotions is fundamentally different from modeling conventional motion styles such as strutting, zombie-like, etc. These conventional styles can be considered ``discrete'', such that clips or images can be categorized as belonging to a particular style. Emotions, on the other hand, span a continuous space~\cite{vad}, such that different motion clips can have different intensities of the same emotion. For example, a somewhat happy gait is expressed differently from one that is extremely happy. For conventional styles, it is generally not needed to account for such intensities, e.g., slightly strutting vs. heavily strutting. To account for this continuous nature, we assume each gait in the input dataset is associated with an emotion vector whose components are the $C$ categorical emotion terms, \textit{i.e.}, each emotion $m$ is a vector in $\mathbb{R}^C$. In practice, the value of each element $l$ in an emotion vector $m$ is the relative count of the number of annotators who labeled the corresponding gait with the categorical emotion term $l$. Also, for training, and due to the practical limitations of separately annotating the emotion at each time step, we repeat the same annotated emotion $m$ in all the time steps. In other words, we assume the emotion vector remains unchanged throughout the corresponding input gait.

\subsubsection{Pose Affective Features}\label{subsubsec:pose_aff_features}
Prior studies in psychology have shown that various physically-based pose features observed per-frame during a gait, better known as \textit{affective features}, aid the identification of perceived emotions from gaits~\cite{karg2013body,crenn2016body}. Roether et al.~\cite{critical_gait_features} identified such a set of necessary pose affective features for human perception. To make these features suitable for machine perception, prior works such as~\cite{tanmay_emotions,step,taew} have come up with necessary sets of scale-independent pose affective features that can be computed geometrically. Scale independence is an important factor in such intra-frame affective features, as observers can identify emotions irrespective of the distance from or the physical stature of the subject. In our work, we use the following three types of scale-independent pose affective features to encode the relevant emotion information:

\paragraph{Angles.} We use the angles subtended by a pair of joints at a third joint. For example, the angle between the two shoulder joints at the neck measures slouching, an indicator of valence and arousal.
\paragraph{Distance ratios.} We use the ratios of the distances between two pairs of joints. For example, the ratio of the distance between the two feet joints to the distance between the neck and the root joints measures the stride, which can indicate arousal and dominance.
\paragraph{Area ratios.} We use the ratios of areas formed by two triplets of joints. These can be considered as amalgamations of the angle- and the distance ratio-based features and they can be used to supplement observations from both these types of features. For example, the ratio of the area of the triangle formed by the hand indices and the neck to the area of the triangle formed by the toes at the root can be used to simultaneously measure arm swings and strides, which can collectively indicate the valence, arousal, and dominance.

We use $11$ angles, four distance ratios and three area ratios for a total of $18$ pose affective features, which we collectively denote as $a^t \in \mathbb{R}^{18}$ at each time step $t$. We list all these pose affective features in Figure~\ref{fig:affective_features}, and direct the interested reader to~\cite{taew} for a detailed analysis on choosing these features.

\begin{figure}[t]
    \centering
    \includegraphics[width=0.7\columnwidth]{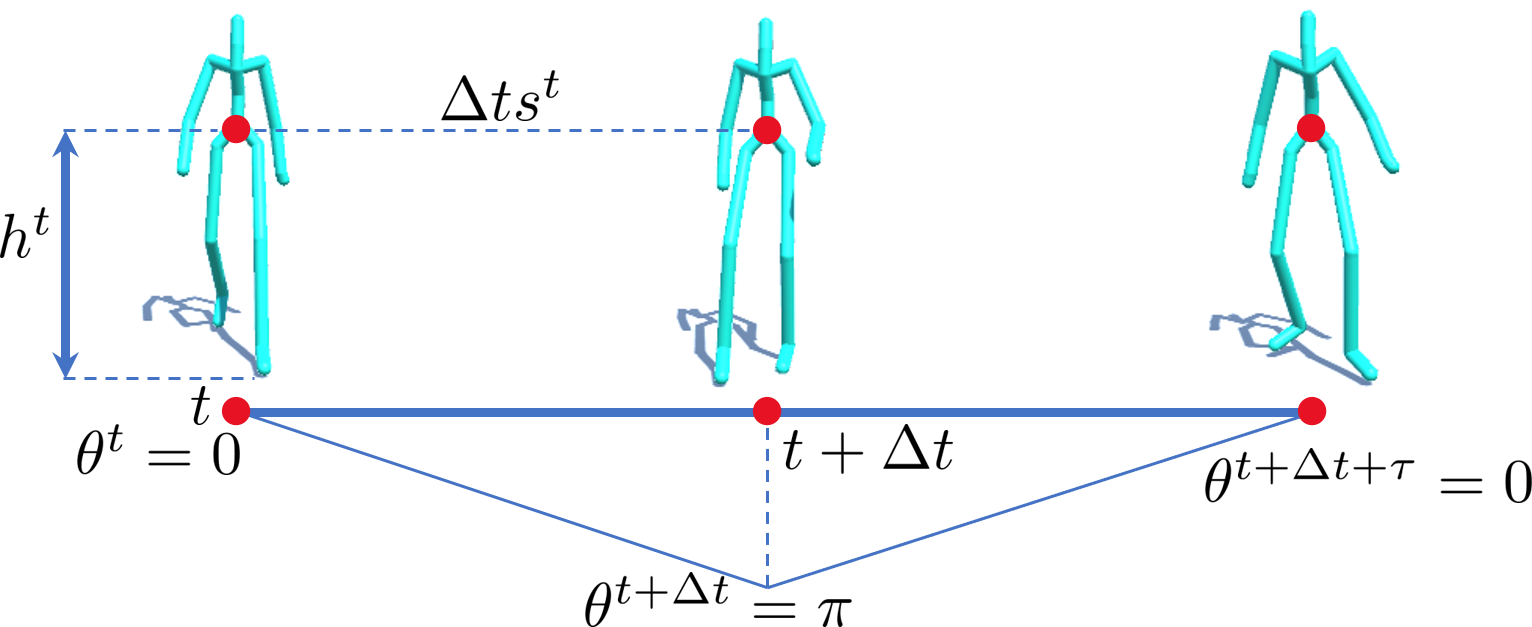}
    \caption{\textbf{Movement features:} We show the root height from the ground $h^t$, the root speed $s^t$, and the stepping phase $\theta^t$. The root speed is the distance travelled between time steps $t$ and $t-1$. The stepping phase $\theta^t=0$ when the left foot touches the ground at time step $t$, $\theta^{t + \Delta t}=\pi$ when the right foot touches the ground at time step $t + \Delta t$, and $\theta^{t + \Delta t + \tau}=0$ when the left foot touches the ground again. We fill in the values for $\theta^t$ between these time steps using linear interpolation. We use these features in our autoregression network.}
    \label{fig:movement_features}
    \vspace{-10pt}
\end{figure}

\subsubsection{Movement Features}\label{subsec:representing_trajectory}
We use a trajectory followed throughout a gait to extract pertinent movement information for our network to generate gaits following given trajectories. We use two kinds of movement features: \textit{root joint features} and \textit{stepping features}. The former consists of root height deviation, root speed (a low-pass filtered component), and root orientation difference trajectory curvature.
The latter consists of stepping phase and foot-step frequency, which are obtained from the trajectory. Figure~\ref{fig:movement_features} illustrates some of these features. Apart from movement information, the root joint features and the foot-step frequency also provide inter-frame or dynamic affective information for emotional expressions. We define these features below.

\paragraph{Root joint features} The root height deviation $\parens{h^t}$ is the signed difference of the height of the root joint from its mean height from the ground plane across the time steps. Subjects expressing emotions with higher arousal tend to have their upper bodies more upright, thus keeping the root height above the mean more often than subjects expressing emotions with lower arousal. In our case, the $XZ$ plane is the ground plane, making the root height $h^t = y_0^t$.
    
The root speed $\parens{s^t}$ is the magnitude of the difference of the 2D position of the root joint, as projected on the ground plane, between the current step $t$ and the previous step $t-1$. Root speed helps indicate the arousal as well, with higher arousal tending to result in faster speeds more often. We represent the root speed as $s^t = \norm{\bracks{x_0^t, z_0^t}^\top - \bracks{x_0^{t-1}, z_0^{t-1}}^\top}$. With root speed, we also use its loss-pass filtered component $\Bar{s}^t$. This reduces the high-frequency noise in the root speed, which is especially useful when the network learns on trajectories with high curvatures.
    
The root orientation difference $\parens{\delta^t}$ is the angular difference between the root orientation $\alpha^t$ w.r.t. the world coordinates and the tangent $\tau^t$ to the 2D root joint positions $\bracks{x_0^t, z_0^t}^\top$ on the ground plane, w.r.t. the world coordinates at each time step $t$. We express the tangent using forward difference, \textit{i.e.}, $\tau^t = \bracks{x_0^t, z_0^t}^\top - \bracks{x_0^{t-1}, z_0^{t-1}}^\top$. Then we have $\delta^t = d_\textrm{ang}\parens{\bracks{\sin\alpha^t, \cos\alpha^t}^\top, \tau^t/\norm{\tau^t}}$, where $d_\textrm{ang}$ denotes the unsigned smaller angle between the unit vectors.

\paragraph{Stepping features} The trajectory curvature $\parens{\kappa^t}$ is the norm of the second-order derivative of the 2D positions of the root joint on the ground plane, or equivalently, the derivative of the root joint tangents $\tau^t$ on the ground plane. We compute this using forward difference as well, \textit{i.e.} $\kappa^t = \norm{\tau^t - \tau^{t-1}}$.

The stepping phase $\parens{\theta}$ represents the phases of the feet between the time steps where they touch the ground. We consider a half-period to be the time from the instant of one foot touching the ground, to the subsequent instant when the other foot touches the ground. Given the half-periods, we define the stepping phase $\theta^t$ at each time step $t$ as follows. We assign a phase $\theta^t=0$ when the left foot touches the ground and a phase $\theta^t=\pi$ when the right foot touches the ground, filling in the intermediate phases through linear interpolation.

The foot-step frequency $\parens{\omega^t}$ is the angular velocity of the foot joints. Apart from generating realistic walk cycles (the motion between ipsilateral footsteps), this feature also supplements the root speed information to indicate the arousal in the emotions expressed by the gait. We compute the foot-step frequency at each time step $t$ as the difference between the phase at that time step and the previous time step $t-1$, \textit{i.e.} $\omega^t = \theta^t - \theta^{t-1}$.

%% Approach
\begin{figure*}[t]
    \centering
    \includegraphics[width=0.8\textwidth]{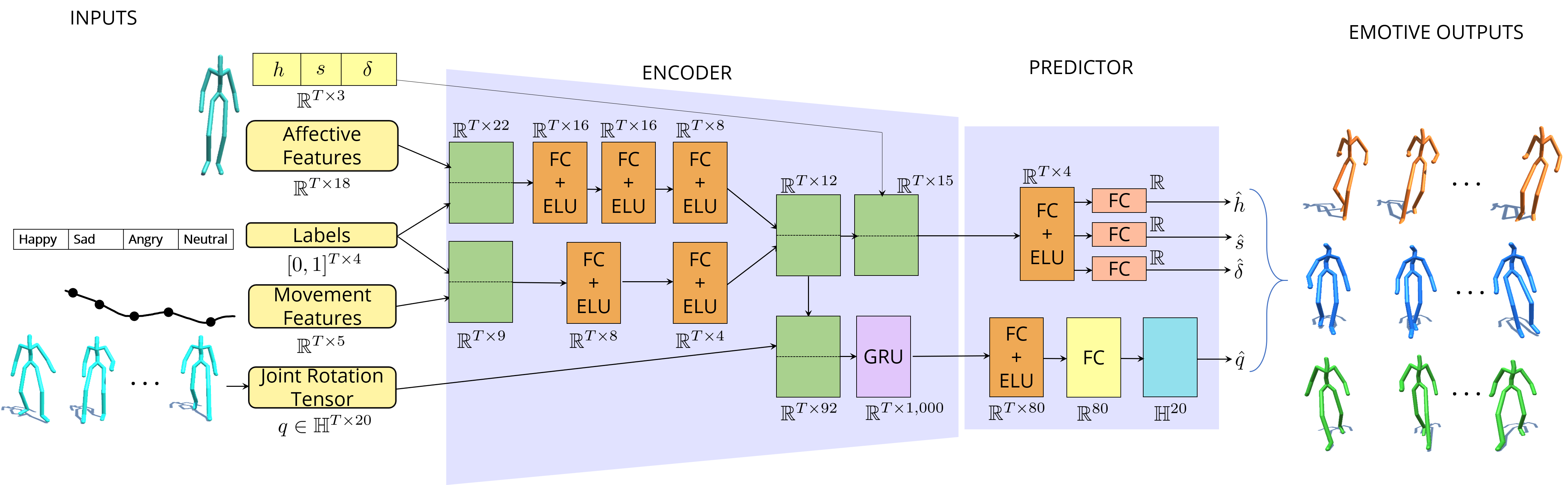}
    \caption{\textbf{Our autoregression network for emotive gaits:} Our network takes in the joint rotations, input emotions as vectors consisting of probabilities for happy, sad, angry, and neutral, pose affective features, and movement features and jointly maps them to a latent representation space through the encoder. The predictor then takes in the latent representations and predicts gaits for subsequent time steps that follow the input trajectory while expressing the input emotions. The green boxes denote concatenation, and the cyan box at the end of the predictor denotes normalization of the variables to versors.}
    \label{fig:network}
    \vspace{-10pt}
\end{figure*}

\section{Autoregression}\label{sec:approach}

Given a sequence of values with some information content, the overall goal of autoregression is to predict subsequent values in the sequence to maintain a similar information content~\cite{sl_survey}. In our work, we use an autoregression network to encode gaits with given emotions and predict subsequent gaits while maintaining the same emotions.
Our network consists of an encoder followed by a predictor. The encoder takes in the joint positions and rotations of the input gait for a number of time steps and extracts pose affective features and root joint features from the input gaits. Next, it learns the encoding functions to jointly map these extracted features, the corresponding input emotions, and the stepping features (such as curvature and foot-step frequency) to latent representations.

The predictor learns to compute the inverse mapping from the latent representations to the necessary affective and trajectory features and, by extension, the joint positions, and rotations, in the subsequent time steps.

We train the encoder and the predictor in tandem, by adding the predictor's output back to the encoder's input and advancing the temporal window of the encoder. We are able to achieve emotional expressiveness by training our network with input gaits for different emotions and forcing the network to learn to predict the corresponding pose affective features in the prediction time steps from its latent representations. We simultaneously enforce a robust constraint on the network to adapt its trajectory features such that its predicted movements are close to the corresponding movements in the ground truth. This enables the network to take sharp turns and follow bends in the trajectory without smoothing out feet movements.
% DO YOU PRESENT DETAILS OF THE TRAINING LATER (WHAT DATABASE, WHAT LABELS, WHAT SPECIFIC PACKAGE, ETC.)
Our overall approach is shown in Figure~\ref{fig:network}. We now elaborate on the operations of the encoder and the predictor.

\subsection{Encoder}\label{subsec:encoder}
% CAN YOU TALK ABOUT SOME BROAD ISSUES IN THE DESIGN OF THE ENCODER?
In the encoder, we separately combine our emotion vectors $m$ with the pose affective features $a^t$, the stepping features consisting of $\bracks{\sin\theta^t, \cos\theta^t}$ and $\omega^t$, and the root joint features $\Bar{s}^t$ and $\kappa^t$, giving us input vectors $i_1^t = \bracks{a^t, m}^\top \in \mathbb{R}^{18 + C}$ and $i_2^t = \big[\sin\theta^t, \cos\theta^t, \omega^t, \allowbreak \Bar{s}^t, \allowbreak \kappa^t, m\big]^\top \in \mathbb{R}^{5 + C}$.

We pass each of these inputs through a set of $3$ fully connected layers, respectively, collectively denoted as the functions $\textrm{FC}_{\textrm{enc}_1}\parens{\cdot, \phi_{\textrm{FC}_{\textrm{enc}_1}}}: \mathbb{R}^{T \times \parens{18 + C}} \rightarrow \mathbb{R}^{T \times H_1}$ and $\textrm{FC}_{\textrm{enc}_2}\parens{\cdot, \phi_{\textrm{FC}_{\textrm{enc}_2}}}: \mathbb{R}^{T \times \parens{5 + C}} \rightarrow \mathbb{R}^{T \times H_2}$. Here, $H_1$ and $H_2$ denote the number of hidden units in the last fully connected layer in $\textrm{FC}_{\textrm{enc}_1}$ and $\textrm{FC}_{\textrm{enc}_2}$ respectively, and $\phi_{\textrm{FC}_{\textrm{enc}_1}}$ and $\phi_{\textrm{FC}_{\textrm{enc}_2}}$ denote the set of trainable parameters in the two sets of fully connected layers, respectively.

We combine the outputs of these fully connected networks to obtain intermediate representations $\gamma^t$, \textit{i.e.} we have
\begin{equation}
    \gamma^t = \bracks{
    \textrm{FC}_{\textrm{enc}_1}\parens{i_1, \phi_{\textrm{FC}_{\textrm{enc}_1}}}; \textrm{FC}_{\textrm{enc}_2}\parens{i_2, \phi_{\textrm{FC}_{\textrm{enc}_2}}}}
\end{equation}
where $i_1 = \bracks{i_1^0, \dots, i_1^{T-1}}^\top$ and $i_2 = \bracks{i_2^0, \dots, i_2^{T-1}}^\top$.

We then append $\gamma^t$ separately to our input joint rotations $q^t$ and to the remaining root joint trajectory features, $h^t$, $s^t$ and $\delta^t$. We pass the appended rotation data through a GRU to obtain the latent representations $\Tilde{q}$, \textit{i.e.}, we have
\begin{equation}
    \Tilde{q} = \textrm{GRU}_\textrm{versors}\parens{\bracks{q; \gamma}, \phi_{\textrm{GRU}_\textrm{versors}}}
\end{equation}
where
\begin{itemize}
    \item[--] $q = \bracks{q^0, \dots, q^{T-1}}^\top$,
    \item[--] $\gamma = \bracks{i_1^0, \dots, \gamma^{T-1}}^\top$,
    \item[--] $\textrm{GRU}_\textrm{versors}: \mathbb{R}^{T \times \parens{4\parens{J-1} + H_1 + H_2}} \rightarrow \mathbb{R}^{T \times H_3}$,
    \item[--] $H_3$ is the number of hidden units in the final layer of the GRU, and
    \item[--] $\phi_{\textrm{GRU}_\textrm{versors}}$ denotes the trainable parameters in the GRU. 
\end{itemize}

We also pass the appended root joint trajectory features through a fully connected layer $\textrm{FC}_{\textrm{root}}: \mathbb{R}^{T \times \parens{3 + H_1 + H_2} \rightarrow T \times H_4}$, $H_4$ being the number of hidden units in the layer, to obtain the latent representations $\Tilde{h}$, $\Tilde{s}$,
and $\Tilde{\delta}$. That is, we have
\begin{equation}
    \bracks{\Tilde{h}, \Tilde{s}, \Tilde{\delta}}^\top = \textrm{FC}_{\textrm{root}}\parens{\bracks{h; s; \delta; \gamma}, \phi_{\textrm{FC}_{\textrm{root}}}}
\end{equation}
where
\begin{itemize}
    \item[--] \resizebox{0.9\columnwidth}{!}{$h = \bracks{h^0, \dots, h^{T-1}}^\top$, $s = \bracks{s^0, \dots, s^{T-1}}^\top$, $\delta = \bracks{\delta^0, \dots, \delta^{T-1}}^\top$},
    \item[--] $\phi_{\textrm{FC}_{\textrm{root}}}$ denotes the trainable parameters in the fully connected layer.
\end{itemize}

\subsection{Predictor}\label{subsec:predictor}
Our predictor takes in the latent representations of the joint rotations and the root joint trajectory features from the encoder and learns to predict the same for $T_\textrm{pred}$ subsequent time steps such that the corresponding generated gaits follow the input trajectory while expressing the input emotion vectors. The predictor consists of a set of $2$ fully connected layers, denoted as $\textrm{FC}_\textrm{versors} : \mathbb{R}^{T \times H_3} \rightarrow \mathbb{R}^{T_\textrm{pred} \times 4\parens{J-1}}$ to predict the joint rotations, and three separate fully connected layers, $\textrm{FC}_h : \mathbb{R}^{T \times H_4} \rightarrow \mathbb{R}^{T_\textrm{pred}}$, $\textrm{FC}_s : \mathbb{R}^{T \times H_4} \rightarrow \mathbb{R}^{T_\textrm{pred}}$, and $\textrm{FC}_\delta : \mathbb{R}^{T \times H_4} \rightarrow \mathbb{R}^{T_\textrm{pred}}$ to compute the respective root joint features. Thus, we have,
\begin{align}
    \hat{q} &= \textrm{FC}_\textrm{versors}\parens{\Tilde{q}, \phi_{\textrm{FC}_\textrm{versors}}}, \\
    \hat{h} &= \textrm{FC}_h\parens{\Tilde{h}, \phi_{\textrm{FC}_h}}, \\
    \hat{s} &= \textrm{FC}_s\parens{\Tilde{s}, \phi_{\textrm{FC}_s}}, \\
    \hat{\delta} &= \textrm{FC}_\delta\parens{\Tilde{\delta}, \phi_{\textrm{FC}_\delta}}
\end{align}
where, as usual, $\phi_{\textrm{FC}_\textrm{versors}}$, $\phi_{\textrm{FC}_h}$, $\phi_{\textrm{FC}_s}$, and $\phi_{\textrm{FC}_\delta}$ respectively denote the respective trainable parameters.

From the predicted joint rotations and root joint features at each time step $t$, we can also compute the predicted pose $\hat{X}^t$ and the corresponding pose affective features $\hat{a}^t$. We use these predicted variables, together with the input data, to train our network according to a curriculum schedule described in Section~\ref{subsec:training}.

\subsection{Loss Function for Training}\label{subsec:loss}
We now describe the formulation of the loss function for training and validating our network. The loss function should accurately constrain the network both to learn the emotional expressions in the input gaits as well as to follow the gaits' trajectories in the subsequent time steps with plausible joint motions. We capture all these requirements in the loss function using four loss terms: the motion loss, the pose loss, the pose affective features loss, and the root joint features loss.

\subsubsection{Motion loss $\parens{\mathcal{L}_\textrm{motion}}$}
This loss ensures that the predicted joint motions remain plausible, \textit{i.e.} close to the ground truth joint motions. To compute this loss, we measure the angle difference between the ground truth rotations $q^t$ and the predicted rotations $\hat{q}^t$ on each joint at each prediction time step $t$. We also add the unit norm constraint on the predicted versors as regularization. Thus, we write the motion loss as
\begin{equation}
    \resizebox{0.9\columnwidth}{!}{$
    \mathcal{L}_\textrm{motion} := \sum_{j, t} \norm{\textrm{q2e}\parens{q_j^t} - \textrm{q2e}\parens{\hat{q}_j^t}}^2 + \lambda_\textrm{versor}\parens{\norm{\hat{q}_j^t} - 1}^2
    $}
    \label{eq:motion_loss}
\end{equation}
where $\textrm{q2e}:\mathbb{H} \rightarrow \bracks{0, 2\pi}^3$ maps the versors to corresponding Euler angles, and the summation is over all the joints across all the prediction time steps.

\begin{table}[t]
    \centering
    \caption{\textbf{Joint Position and Rotation Errors.} We compute position error relative to the longest diagonal of the bounding box of the characters we test, and we compute rotation errors in degrees. The performance of our method is on par with the current state-of-the-art in motion generation.}
    \label{tab:errors}
    \begin{tabular}{lcc} 
        \toprule
        \textbf{Method\Tstrut} & \textbf{Pose Error} & \textbf{Rotation Error}\\
        \midrule
        PFNN~\cite{pfnn} & 0.19 & 0.06 \\
        \midrule
        QuaterNet~\cite{quaternet} & 0.16 & 0.05 \\
        \midrule
        Emotive Gait Styling & 0.12 & 0.04 \\
        \bottomrule
    \end{tabular}
    \vspace{-10pt}
\end{table}

\subsubsection{Pose loss $\parens{\mathcal{L}_\textrm{pose}}$}
The pose loss supplements the motion loss by adding an extra regularization to maintain plausible predicted joint motions. We require the predicted character poses $\hat{X}^t$ at each prediction time step, obtained using the predicted versors $\hat{q}_j^t$, to be as close as possible to the corresponding ground truth poses $X^t$. However, we do not require our predicted poses to follow the same trajectory as the ground truth poses since the desired trajectory will be provided to us at test time. We, therefore, subtract the root joint position from all the other joints at every time step and write our pose reconstruction loss $\mathcal{L}_\textrm{pose}$ as
\begin{equation}
    \mathcal{L}_\textrm{pose} := \sum_t\sum_{j=1}^{J-1} \norm{\parens{X_j^t - X_0^t} - \parens{\hat{X}_j^t - \hat{X}_0^t}}^2.
    \label{eq:pose_loss}
\end{equation}

\subsubsection{Pose affective features loss $\parens{\mathcal{L}_\textrm{aff}}$}
This loss constrains the network to predict pose affective features similar to the ones computed from the input gaits. Therefore, it forces the network to maintain the emotional expressions in the gaits. We compute this loss by measuring the norm difference between the ground truth affective features $a^t$ and the predicted features $\hat{a}^t$. We write it as
\begin{equation}
    \mathcal{L}_\textrm{aff} = \sum_t\norm{a^t - \hat{a}^t}^2.
    \label{eq:aff_loss}
\end{equation}

\subsubsection{Root joint features loss $\parens{\mathcal{L}_\textrm{root}}$}
This is a robust loss that we use to constrain the network to follow the ground truth gait trajectory at the prediction trajectory. The robustness ensures that the prediction follows sharp turns and bends in the trajectory without smoothing out the foot joint movements. We compute this loss by measuring the $L_1$ norm difference between the ground truth root joint features $h$, $s$, and $\delta$, and the predicted features $\hat{h}$, $\hat{s}$, and $\hat{\delta}$ given by our network. We write it as
\begin{equation}
    \mathcal{L}_\textrm{root} = \sum_t\norm{\bracks{h^t; s^t; \delta^t} - \bracks{\hat{h}^t; \hat{s}^t; \hat{\delta}^t}}_1.
    \label{eq:root_loss}
\end{equation}

\subsubsection{Foot contact loss $\parens{\mathcal{L}_\textrm{ft\_ct}}$}
We also require the generated characters to walk naturally without any foot sliding. Therefore, we add a robust $L_1$ norm loss to constrain the heel and toe positions of the generated character to match the ground truth heel and toe positions. The robustness ensures that the prediction follows sharp turns and bends in the trajectory without smoothing out the foot joint movements. We write this loss as
\begin{equation}
    \mathcal{L}_\textrm{ft\_ct} = \sum_t\norm{\bracks{lh^t; lt^t; rh^t; rt^t} - \bracks{\hat{lh}^t; \hat{lt}^t; \hat{rh}^t; \hat{rt}^t}}_1.
    \label{eq:ftct_loss}
\end{equation}

Finally, we linearly combine all these loss terms to formulate our overall loss function $\mathcal{L}$, which we write as
\begin{equation}
    \resizebox{0.9\columnwidth}{!}{$
    \mathcal{L} = \lambda_\textrm{motion}\mathcal{L}_\textrm{motion} + \lambda_\textrm{pose}\mathcal{L}_\textrm{pose} + \lambda_\textrm{aff}\mathcal{L}_\textrm{aff} +
    \lambda_\textrm{root}\mathcal{L}_\textrm{root} +
    \lambda_\textrm{ft\_ct}\mathcal{L}_\textrm{ft\_ct}
    $}
    \label{eq:training_loss}
\end{equation}
where $\lambda_\textrm{motion}$, $\lambda_\textrm{pose}$, $\lambda_\textrm{aff}$, $\lambda_\textrm{root}$ and $\lambda_\textrm{ft\_ct}$ are the corresponding scaling terms and assign relative importance to the different loss terms.

%% Results
\section{Results}\label{sec:results}
We show the performance of our autoregression network on the training dataset described below. We briefly describe the training dataset in Section~\ref{subsec:dataset} and discuss our augmented dataset in Section~\ref{subsec:dataset_augmented}. We report our training routine in Section~\ref{subsec:training}, elaborate on our performance benchmarks in Sections~\ref{subsec:experiments} and \ref{subsec:comparisons}, and discuss the contributions of our novel components through ablation studies in Section~\ref{subsec:ablation}. We summarize the details of integrating our setup with the AR environment in Section~\ref{subsec:integration_with_ar}. For a video demonstration of the results, please refer to our supplementary material.

\begin{figure}[t]
    \centering
    \includegraphics[width=0.8\columnwidth]{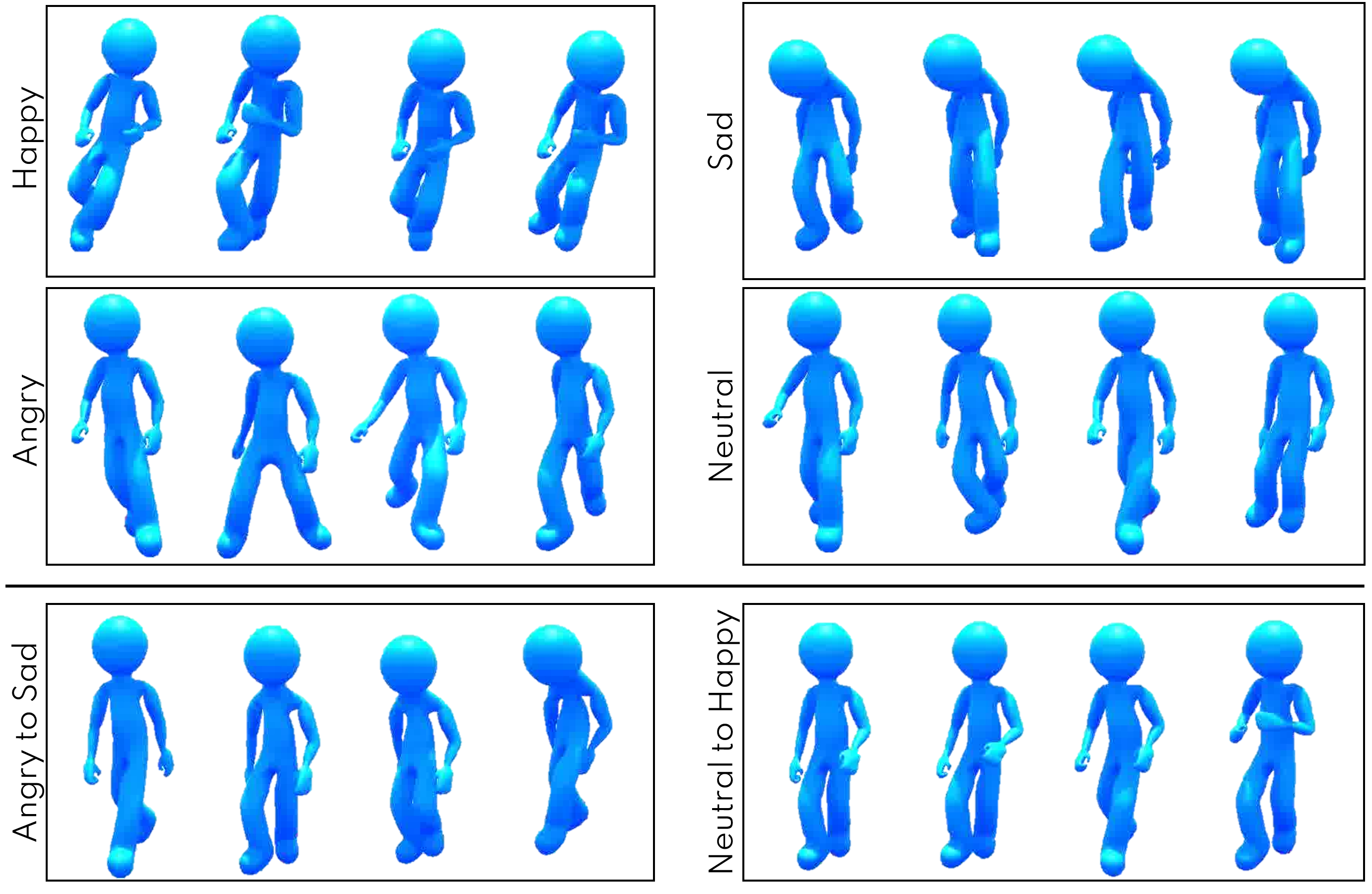}
    \caption{\textbf{Emotional expressions and transitions.} Each row shows four snapshots of synthesized gaits in temporal sequence from left to right. The top two rows show gaits with single emotions. The bottom row shows gaits transitioning from one emotion to another.}
    \label{fig:emotion_expression_transition}
    \vspace{-10pt}
\end{figure}

\subsection{Dataset for Training}\label{subsec:dataset}
The datasets we use consist of temporal 3D pose sequences of human gaits for different types of walking, running, and other locomotion activities. This dataset was collected from various 3D pose sequence datasets, including BML~\cite{bml}, Human3.6M~\cite{human3.6m}, ICT~\cite{ict}, CMU-MoCap~\cite{cmu_mocap}, ELMD~\cite{elmd}, and Emotion-Gait dataset~\cite{step}. All gaits in the dataset are $240$ frames long and playable at $10$ fps. Due to memory constraints, we sampled every $\raisedth{4}$ frame and used the resultant $60$ frames as input data to our network, \textit{i.e.} we had $T = 60$ for all our data points. In total, the dataset consists of $1,835$ gaits with corresponding emotion vectors available.

We used $80\%$ of our gait dataset for training our network, $10\%$ for validation, and kept the remaining $10\%$ of the dataset for testing the emotional-expressiveness and trajectory-following performances. We performed this split randomly, and the network never sees the validation and the test sets during training.

\subsection{Augmented Dataset: Synthesized Emotive Gaits}\label{subsec:dataset_augmented}
At test time, our network is able to generate predicted gaits on trajectories it did not encounter during training. Our network is also able to transition between different emotions on the test gaits as a result of the learned inverse mapping from the latent representation space of the encoder. We, therefore, use our network to augment synthesized gaits to the Emotion-Gait benchmark datasets. We generate gaits on $20$ trajectories not present in the dataset, with $100$ emotions, also not present in the dataset, on each trajectory, for a total of $2,000$ new gaits. We also perform transitions between $50$ pairs of emotions on each trajectory, picking a pair of emotions from the $100$ novel ones without replacement, thus adding another $1,000$ new gaits, taking the total new gaits added to $5,000$.
\begin{figure}[t]
    \centering
    \includegraphics[width=0.8\columnwidth]{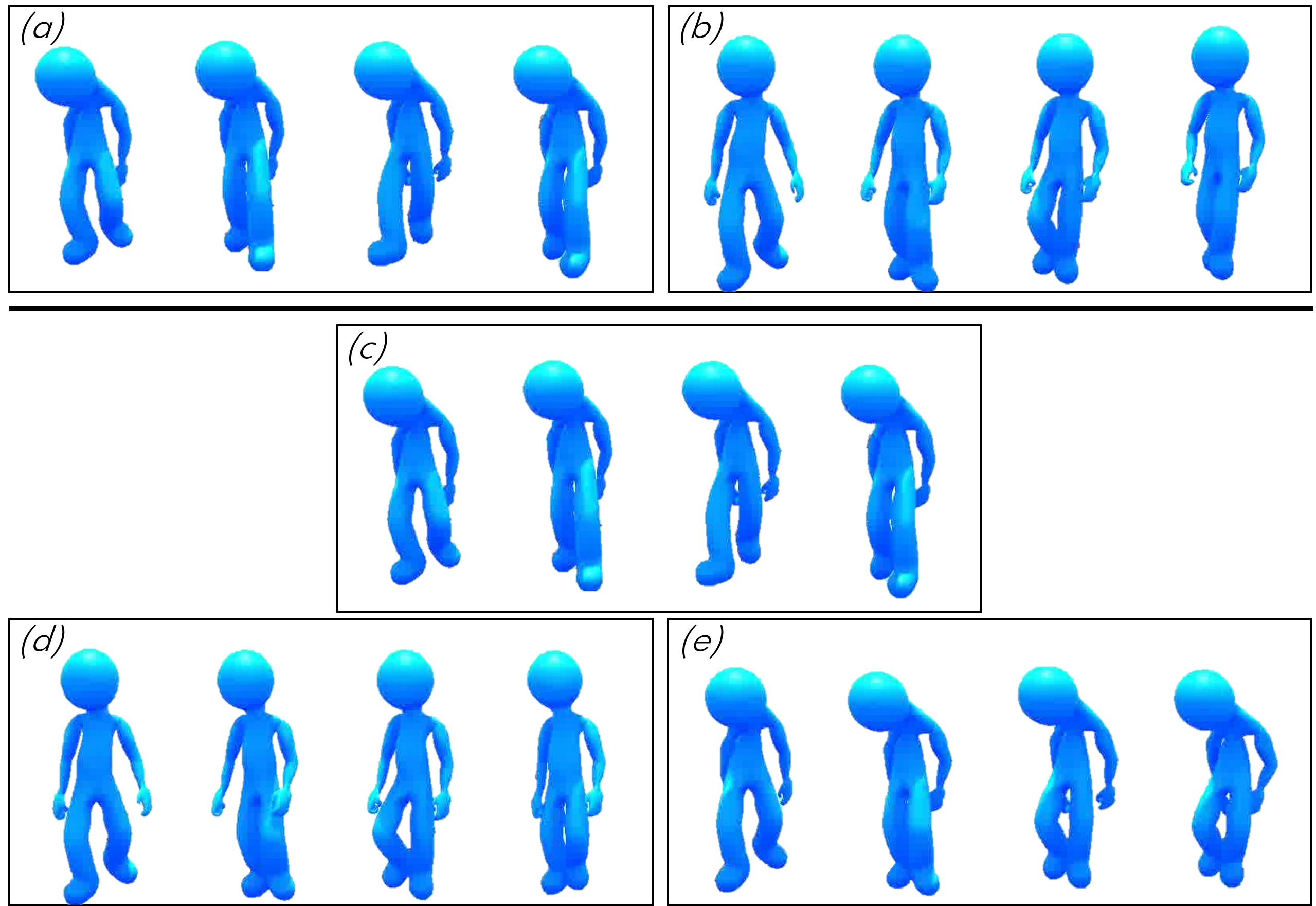}
    \caption{\textbf{Comparison and Ablation Studies.} $\parens{a}$ and $\parens{c}$ shows emotive four snapshots in temporal sequence from left to right gaits generated by our network following user-driven trajectories. $\parens{b}$ shows the results at the same four time instances for QuaterNet, which has no emotive component.
    $\parens{d}$ shows the results at the same four time instances for our network without the affective feature component. In this case, the gait is able to follow the trajectory, but not express the emotions (\textit{e.g.}, no shoulder slouching to indicate sadness).
    $\parens{e}$ shows the results at the same four time instances for our network without the movement feature component. In this case, the gait is able to express emotions, but not follow the desired trajectory.}
    \label{fig:emotion_comparison}
    \vspace{-10pt}
\end{figure}

\subsection{Training Routine}~\label{subsec:training}
We use the Adam optimizer~\cite{adam} with a learning rate of $0.001$, which we decay by a factor of $0.999$ at every epoch. We use the ELU activation~\cite{elu} on all the fully connected layers in the network.

Similar to Pavllo et al.~\cite{quaternet}, we use a curriculum scheduling technique~\cite{curriculum_schedule} to train our network. We begin training by presenting various sequences of the input data and control features, each of length $T$, to our network, and the network predicts the rotations and translations for a single subsequent time step. This is equivalent to having a teacher forcing ratio of $1$. At every subsequent epoch $E$, we decay the teacher forcing ratio by $\beta = 0.995$, \textit{i.e.} with probability $\beta^E$, we supplement the data and controls at each input time step with the network's predicted data and controls at that time step. In other words, we progressively expose the network to more and more of its own predictions to make further predictions. Curriculum scheduling thus helps the network gently transition from a teacher-guided prediction routine to a self-guided prediction routine, which significantly speeds up the training process.

\begin{figure}[t]
    \centering
    \includegraphics[width=0.7\columnwidth]{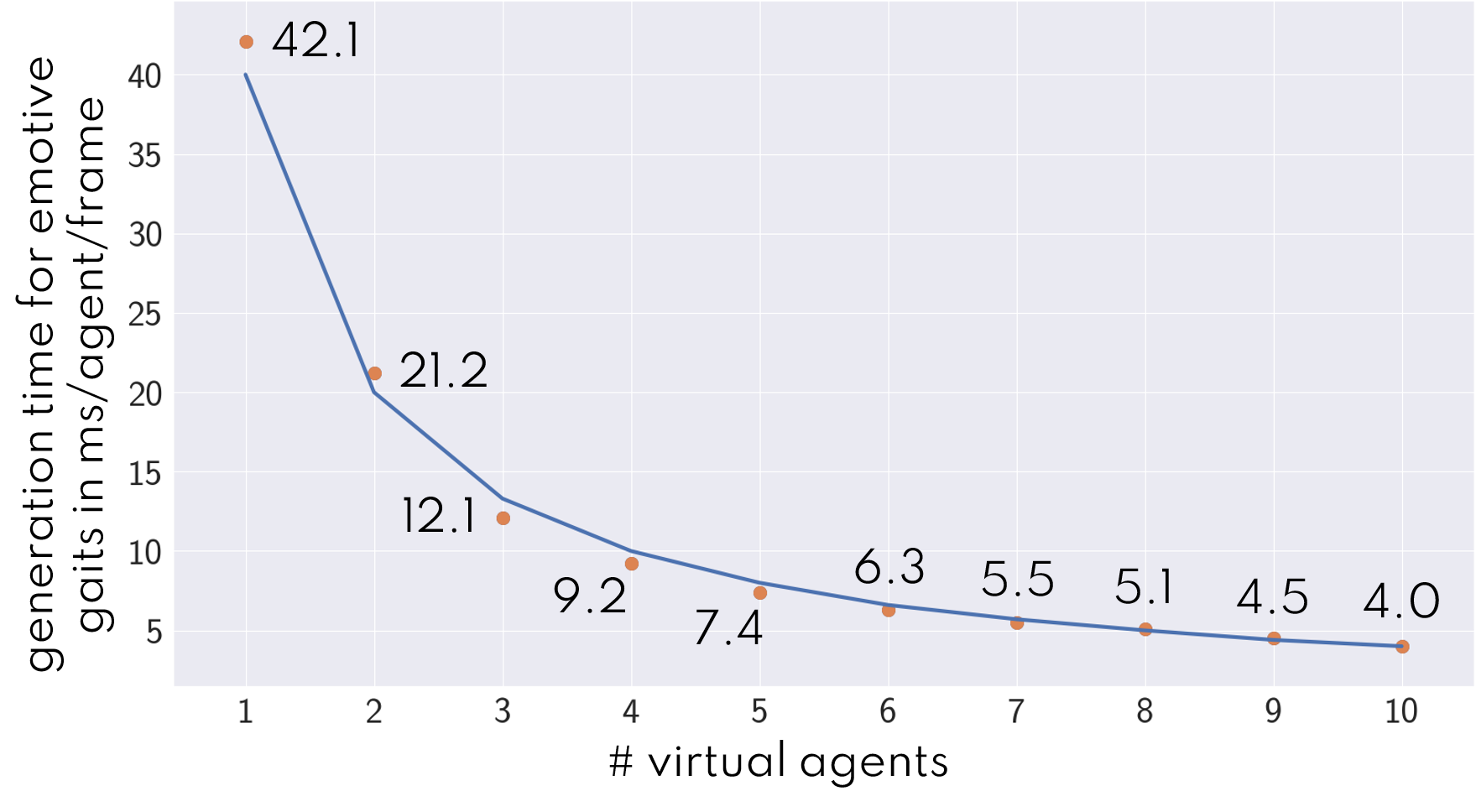}
    \caption{\textbf{Time to generate emotive gaits:} We highlight the generation and rendering time for emotive gaits on a pair of GPUs. The average time per agent per frame decreases, as we increase the number of virtual agents.}
    \label{fig:render_times}
    \vspace{-10pt}
\end{figure}

\begin{table*}[t]
    \centering
    \caption{\textbf{Likert scales for observed pose affective features.} The Likert scale response categories we provided users for the four broad observed pose affective features. Our goal is to evaluate if different users find the Likert scale values of these features similar for the same emotion, which would indicate these features are relevant for perceiving the emotions.}
    \label{tab:user_study_likert}
    \resizebox{0.8\linewidth}{!}{
    \begin{tabular}{lC{2.8cm}C{2.8cm}C{2.8cm}C{2.8cm}C{2.8cm}}
        \toprule
        Feature & \multicolumn{4}{c}{Likert Scale Response Categories} \\
        \cmidrule{2-6}
        & Value = 0 & Value = 1 & Value = 2 & Value = 3 & Value = 4 \\
        \midrule
        Torso & Contracted, bowed & Somewhat contracted & Neither contracted nor expanded & Somewhat expanded & Expanded, stretched \\
        \midrule
        Arms & Contracted, close to the body & Somewhat contracted & Neither contracted nor expanded & Somewhat expanded& Expanded, away from the body \\
        \midrule
        Gait Pace & Sustained, leisurely, slow & Somewhat sustained & Neither sustained nor hurried & Somewhat hurried & Hurried, sudden, fast \\
        \midrule
        Gait Flow & Free, relaxed, uncontrolled & Somewhat free & Neither free nor bound & Somewhat bound & Bound, tense, controlled \\
        \bottomrule
    \end{tabular}
    }
    \vspace{-10pt}
\end{table*}

We train our network for $500$ epochs, which takes around $18$ hours on an Nvidia GeForce GTX 1080Ti GPU with $12$ GB memory. We use $90\%$ of the available data for training our network, and validate its performance on the remaining $10\%$ of the data. We also observed that our network performs well for any values of the scaling terms in Equation~\ref{eq:training_loss} between $0.5$ and $2.0$. We used a value of $1.0$ for all of the performances reported in our experiments in Section~\ref{subsec:experiments}.

\subsection{Performance Benchmarks}\label{subsec:experiments}
We note here that methods generating motions with discrete styles test generalizability by providing their network with novel discrete-style labels not seen during training~\cite{few_shot_homogeneous,unpaired_style_transfer}, as opposed to ranges in the styles. Our network, on the other hand, learns the mapping between gaits and the underlying continuous space of emotions, rather than the mapping between the gaits and annotated emotions in the training samples. As a result, we test the generalizability of our network by generating gaits corresponding to continuous ranges of emotion vectors not seen during training. In order to benchmark our network on its ability to generalize to these continuous ranges of emotions, we perform experiments on emotion expressiveness and emotion transitions.

\subsubsection{Emotion Expressiveness}\label{subsubsec:emotion_expressiveness}
We randomly pick gaits from the test set and extract the first $18$ frames ($\frac{1}{3}^\textrm{rd}$ of the total data length) to provide as inputs to our network. We set the associated emotion vector of the input gait as the desired emotion, initially set a straight line as the desired trajectory, and predict for $200$ time steps. Figure~\ref{fig:emotion_expression_transition} (top row) shows some snapshots of the results of this evaluation. Next, we evaluate our method on trajectories with bends and sharp turns for $200$ steps (Figure~\ref{fig:emotion_expression_transition}, middle row). We note that the generated gait maintains the emotion of the input, and follows all the trajectories. For example, the gait slows down while taking sharp turns and adjusts its stride and other joint movements such that the affective features remain similar when walking on a path with bends and sharp turns.

\subsubsection{Emotion transitions}\label{subsubsec:emotion_transition}
In this set of evaluations, we modify the desired emotion at each prediction time step to be different from those in the previous time steps, as well as the emotion vector associated with the input gait. To track the performance of our network, we first choose a particular emotion vector for the final prediction time step. Next, we linearly interpolate the value of each element of the emotion vector at each prediction time step separately, including everything from the input vector to the vector at the final time step. We normalize the vector at each time step to convert the values to a probability distribution that can be passed to our network. We test the results of emotion transition on trajectories with bends and turns for $200$ time steps (Figure~\ref{fig:emotion_expression_transition}, bottom row). We observe that our network is able to smoothly transition between the different emotions, with no sharp limb movement or jarring action at any time step.

\subsection{Comparisons with Motion Generation}\label{subsec:comparisons}
We visually compare the performance of our autoregression network with QuaterNet developed by Pavllo et al.~\cite{quaternet} in Figure~\ref{fig:emotion_comparison} (rows $\parens{a}$ and $\parens{b}$). QuaterNet is a state-of-the-art motion prediction network, and our network builds on the core prediction framework of QuaterNet. Since QuaterNet performs motion prediction but not emotional expressiveness, the gaits are not able to express the different emotions.

We also summarize the mean pose errors relative to the scale of the input data and the joint rotation errors in degrees as they are produced by our method, QuaterNet, as well as prediction networks based on alternative approaches such as the phase-functioned neural network (PFNN)~\cite{pfnn} in Table~\ref{tab:errors}. We keep the desired emotion for our network the same as the input emotion vector to perform a fair comparison. We also require ground truth gaits to be available so the prediction time steps can actually compute these errors. Therefore, we present the first $18$ frames of each data point in the test set as inputs to both the methods and compute their predicted motions on the trajectory of the ground truth data for the remaining $42$ frames. We notice negligible differences between the performances of the two networks, showing that our approach is comparable to the state-of-the-art in motion prediction. However, for predicting motion beyond the $60$ frames in the dataset, we noticed that the motions predicted by QuaterNet eventually reduce to no movement and the character comes to a stop, whereas both PFNN and our method can predict plausible motions for up to $200$ prediction steps.

Thus, while current motion generation methods can produce highly realistic gaits for VAs, our method can additionally produce emotional expressiveness for those gaits. Therefore, our method helps improve the social presence of the VAs in an AR environment, as we observe through user evaluations in Section~\ref{sec:user_eval}.

\subsection{Ablation Studies}\label{subsec:ablation}
We have two main contributions to the design of our autoregression network. First, we provide the pose affective features as part of the input and constrain the predicted pose affective features to remain close to the corresponding ground truth during training through the affective loss $\mathcal{L}_\textrm{aff}$ (Eq.~\ref{eq:aff_loss}). This enables our network to achieve emotional expressiveness and emotion transition on the input data. We, therefore, remove this input component and the corresponding loss function from our network and the training process and compare the results on the experiments described in Sections~\ref{subsubsec:emotion_expressiveness} and \ref{subsubsec:emotion_transition}. We observe (Figure~\ref{fig:emotion_comparison}, rows $\parens{c}$ and $\parens{d}$) that the ablated network does not maintain consistent pose affective features across the prediction time steps for the desired emotion.

Second, our network predicts the root joint features, consisting of the root height, the root speed, and the root orientation difference (as detailed in Section~\ref{subsec:representing_trajectory}) alongside the predictions for the joint rotations. To ensure that the predicted motion follows the desired trajectory, we constrain these predicted root joint features to remain close to the corresponding ground truth during training through the root joint loss function $\mathcal{L}_\textrm{root}$ (Eq.~\ref{eq:root_loss}). To underscore the importance of this loss function, we remove it from our training and perform the experiments described in Section~\ref{subsubsec:emotion_transition} on the ablated network. We notice (Figure~\ref{fig:emotion_comparison}, rows $\parens{c}$ and $\parens{e}$) that the ablated network is not able to follow the desired trajectory. For linear trajectories, the predicted gaits often end up being oriented in arbitrary directions and not facing the direction of motion. For trajectories containing bends and turns, once the predicted gaits deviate from the desired trajectories, the ablated network is not able to reduce the deviations in subsequent prediction time steps.

\subsection{Integration with the AR Environment}\label{subsec:integration_with_ar}
Our generative method can generate animation frames at the interactive rate of $40$ ms per frame for $10$ agents in an AR environment (Figure~\ref{fig:render_times}), \textit{i.e.}, at $4$ ms per agent per frame on average, when utilizing two Nvidia GeForce GTX 1080Ti GPUs. We built the realtime AR demo by rigging the generated skeletons to humanoid meshes, modifying the posed meshes to handle minor visual distortions caused by body shape mismatch, and streaming a virtual environment containing the animated characters to the Microsoft Hololens.
\paragraph{Rigging} Rigging the humanoid meshes to the generated skeletons requires that the rest pose of the generated skeleton is not modified to accommodate the desired mesh, as this would invalidate the rest of the animation. Thus, the rigging process must be done in reverse; the desired meshes must already contain a skeleton that allows us to repose it to meet the rest pose of the generated skeletons. We found free meshes with suitable skeletons from online 3D mesh databases. For the demo, we chose a humanoid stick figure and some humans from the Microsoft Rocketbox collection~\cite{rocketbox}. We performed the rigging in Blender 2.7.
\paragraph{Modifications To Rig} Due to the body shapes of our desired meshes not exactly matching that of the people used to generate the original dataset, we use Blender's sculpt tools to iron out any distortions. In order to make the human meshes seem less synthetic, we used the original face bones in the meshes to create blendshapes such as blinking, breathing (mouth), and breathing (chest), which we activated at regular intervals. These blendshapes represent typical human behaviors independent of bodily animations. Our generated skeletons do not contain facial bones, thus, blendshapes are a good option for animating the face without requiring bones.
\paragraph{AR Implementation} We made the realtime AR demo in Unreal 4.24 due to its strong animation system allowing trivial sharing of animation files between meshes. We created An environment in which we show pairs of animations of specific emotions, human or stick figure, with the meshes approximately walking along the real ground. We used the Unreal Hololens plugin to stream the rendered images directly to the Hololens through the Hololens' Holographic Remoting player, which receives images by listening on a specific IP address. Due to the start position of the user being non-deterministic, we also provide key inputs to reposition the animated characters in front of wherever the user is when the key is pressed. The animations loop in order to make it easier for the user to determine differences in gaits and the stick figure characters are given colored materials matching their emotion.

\begin{table}[t]
    \centering
    \caption{\textbf{2-sample Anderson-Darling test statistics.} Based on the statistics, we are unable to reject the null hypothesis that the intended and perceived emotions of the gaits are samples from the same probability distribution, except for the case of Gait 2.}
    \label{tab:anderson_darling_test}
    \resizebox{0.8\columnwidth}{!}{
    \begin{tabular}{rccc}
        \toprule
        Gait \# & Value of Statistic & $p$-value & Reject Null Hypothesis? \\
        \midrule
        $1$ & $0.311$ & $> 0.25$ & Not able to \\
        \midrule
        $2$ & $2.111$ & $0.04$ & With $96\%$ confidence \\
        \midrule
        $3$ & $0.311$ & $> 0.25$ & Not able to \\
        \midrule
        $4$ & $-0.615$ & $> 0.25$ & Not able to \\
        \midrule
        $5$ & $-0.615$ & $> 0.25$ & Not able to \\
        \midrule
        $6$ & $-0.611$ & $> 0.25$ & Not able to \\
        \midrule
        $7$ & $-0.069$ & $> 0.25$ & Not able to \\
        \midrule
        $8$ & $-1.081$ & $> 0.25$ & Not able to \\
        \bottomrule
    \end{tabular}
    }
    \vspace{-10pt}
\end{table}

\paragraph{Animation artifacts} We observe some jerkiness in the animation of the human characters in AR. The major sources of this jerkiness are (i) jerky motion of the user wearing the HoloLens, (ii) issues in the HoloLens software, e.g., frame rate clipping, and (iii) using textures instead of deformable cloth materials for the low-poly human models, which makes the jerkiness more apparent due to aliasing. We observe much-reduced jerkiness for the textureless stick figures in AR, and almost none when rendering the stick figures in a purely virtual environment with a known ground plane.

%% User Eval
\section{User Evaluation}\label{sec:user_eval}
We conducted a web-based user study with our generated emotive gaits to test the following \textbf{null hypothesis}:

\noindent\textit{The emotion vector used as input to generate each gait, and the emotion vector obtained by taking the arithmetic mean of the emotion vectors perceived by all the users from that generated gait, are two samples of the same statistical distribution.}

In other words, the distribution of emotions we intend for a generated gait is statistically similar to the distribution of emotions perceived by the observing users.
We also obtain the values of the pose affective features observed by the users from the generated gaits on a five-point Likert scale (LS) to validate our choice of pose affective features and emotional expressiveness of the VAs in Section~\ref{subsubsec:pose_aff_features}.

\subsection{Procedure}
The study was divided into three sections. Each section took three to four minutes to complete on average, and the entire study lasted for around ten minutes on average.

In the first section, we showed the users ten-second clips of eight randomly chosen generated gaits, one at a time, and asked them to report the emotion they perceived from each of those gaits. Users could report multiple emotions. For example, if one gait looked less happy to the user than another (but not necessarily sad), then the user could potentially mark that gait as both happy and neutral.

In the second section, we again showed the users ten-second clips of six randomly chosen generated gaits, one at a time. However, in this section, we performed emotion transitions on the generated gaits, so the final emotions were different from the initial ones. We asked the users to report the initial and the final emotions they perceived from these gaits, with the option to report multiple emotions.

In the third section, we showed the users ten-second clips of the same eight generated gaits from the first section, one at a time, and asked them to report the observed values or \textit{intensities} of four broad pose affective features on a five-point LS. The four pose affective features we chose to ask are inspired by the critical features identified in the study by Roether et al.~\cite{critical_gait_features}. We summarize the scales for each of the four features in Table~\ref{tab:user_study_likert}.

\begin{figure}[t]
    \centering
    \includegraphics[width=0.8\columnwidth]{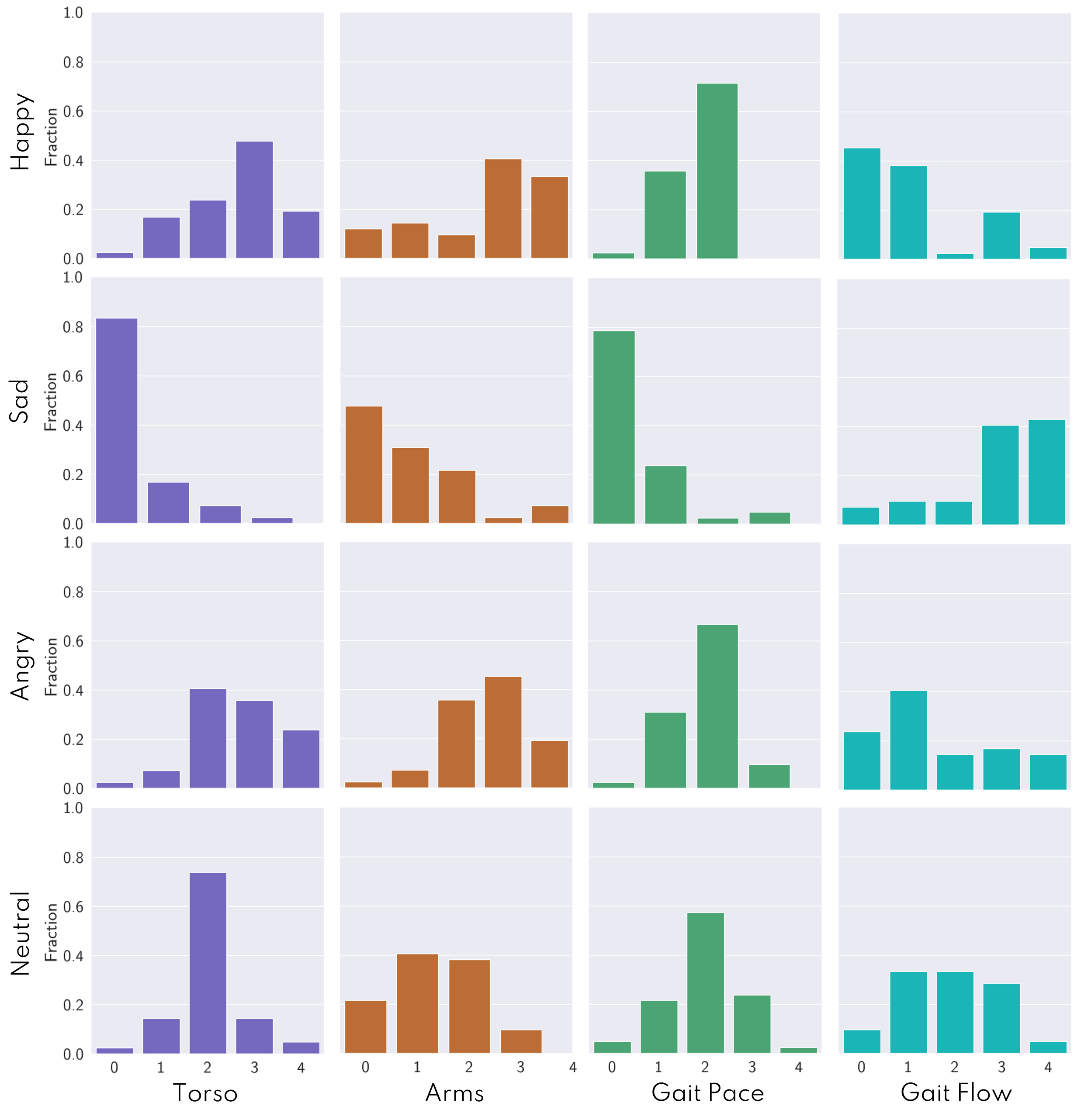}
    \caption{\textbf{Distribution of user votes for the broad pose affective features in gaits from different emotions.} We can observe different distinct modes for the different emotions, indicating that the pose affective features vary between different emotions and are consistent across users for a given emotion. The values in the horizontal axis correspond to the Likert scale values in Table~\ref{tab:user_study_likert}.}
    \label{fig:affective_feature_distribution}
    \vspace{-10pt}
\end{figure}

\begin{figure*}[t]
    \centering
    \includegraphics[width=0.75\textwidth]{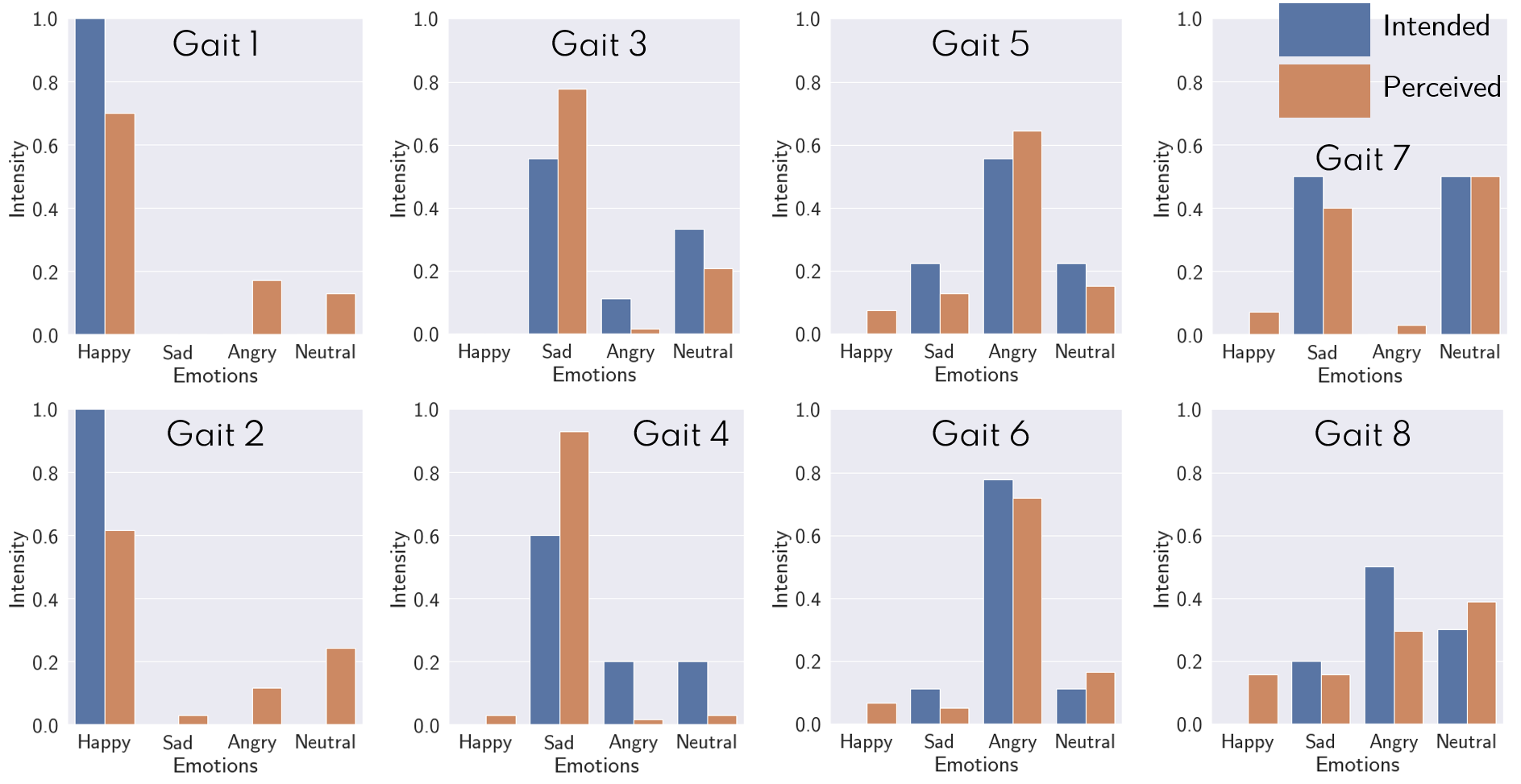}
    \caption{\textbf{Sets of normalized intended and perceived emotion vectors.} As we can observe from the plots and the statistics in Table~\ref{tab:anderson_darling_test}, except for Gait 2, the intended and perceived emotions of the gaits cannot be determined to belong to separate statistical distributions.}
    \label{fig:emotion_distributions}
    \vspace{-10pt}
\end{figure*}

\subsection{Participants}
Since emotion perceptions are influenced by numerous social and cultural factors, we invited participants from diverse demographics to draw useful conclusions.
We had $102$ participants in total, of which $58$ were male and $44$ were female. $31$ male and $26$ female participants were in the age group of $18$-$24$. $25$ male and $14$ female participants were in the age group of $25$-$34$. $2$ male and $4$ female participants were above $35$. Based on the overall test statistics, we did not find any noticeable difference in the emotions perceived from the generated gaits across the different sexes and age groups.

\subsection{Analysis}
We analyze the results on single emotions (section 1), emotion transition (section 2), and pose affective features (section 3). Finally, we report the perceived naturalness of the gaits by the users and miscellaneous analyses in ``other feedback''.

\subsubsection{Single Emotions}
Given the perceived emotions from the first section of the user study, we plot the normalized perceived emotions for eight randomly chosen gaits, as well as the corresponding normalized intended emotions of the generated gaits, in Figure~\ref{fig:emotion_distributions}. The emotions are denoted as four-component vectors as described in Section~\ref{subsubsec:emotions}. We perform $l_1$ normalization so that each component of the emotion vector represents the intensity of the corresponding emotion.

For each gait, we perform the 2-sample Anderson-Darling test~\cite{anderson_darling_ks} on the null hypothesis that the set of probability values of the perceived emotions and the set of probability values of the intended emotions are samples of the same underlying distribution. Table~\ref{tab:anderson_darling_test} summarizes the test statistic and the corresponding $p$-values for each of the eight gaits in Figure~\ref{fig:emotion_distributions}.

As we can observe from Table~\ref{tab:anderson_darling_test}, we cannot reject our null hypothesis for seven of the eight gaits. This suggests strong statistical evidence that the intended and perceived emotions are statistically similar for those seven gaits. In Gait 2, where we reject the null hypothesis, the intended emotion was fully happy, but the observers mainly perceived it as either happy or neutral, indicating that the intensity of happiness did not come across to some of the observers.

\subsubsection{Emotion Transition}
We performed a similar 2-sample Anderson-Darling test~\cite{anderson_darling_ks} for each of the initial and the final intended and perceived emotions, and were unable to reject the null hypothesis in 10 out of the 12 gaits we tested with. This again provides strong statistical evidence that the intended emotions for the gaits and the corresponding perceived emotions are statistically similar.

In one rejected case, the initial emotion was predominantly angry while the final was predominantly happy, but many observers indicated that both the initial and the final emotions were neutral. In the other case, the transition was from predominantly sad to predominantly happy, but many observers reported the gait to be going from sad to neutral. We hypothesize two possibilities for the mismatches:
\begin{itemize}
    \item the intensities of the initial and final emotions did not come across in the generated gaits,
    \item the observers did not expect the gait to transition between extreme emotions such as angry to happy or sad to happy in the ten-second span of the clip, hence opted for choices they found more reasonable.
\end{itemize}

\subsubsection{Pose affective features}
Our goal here is to validate the usefulness of the pose affective features we use to train our network, as well as the emotional expressiveness of the VAs through these pose affective features. However, evaluating the values of angles and ratios is out of scope for a user-study. We, therefore, opted to measure the user-observed LS values or \textit{intensities} of the broad pose affective features that we used to formulate our geometric features. A good test is to verify if the observed intensities of these broad pose affective features are statistically consistent across different users. If this is verified, it justifies
\begin{itemize}
    \item basing the geometric features on these broad features,
    \item the VAs are able to clearly express the different emotions through different LS values of the pose affective features.
\end{itemize}

We show the distribution of the fraction of users that marked each particular intensity of the four broad pose affective features for each of the single intended emotions in our study in Figure~\ref{fig:affective_feature_distribution}. The values in the horizontal axis correspond to the LS values in Table~\ref{tab:user_study_likert}. From this figure, we can observe different distinct modes in the distribution for the different intended emotions. For example, the mode for the torso is at ``contracted, bowed'' (0) for sad, while it is concentrated more around ``somewhat expanded'' (3) and ``expanded, stretched'' (4) for happy. For angry, the users observed it to be less expanded than happy overall, but less than $10\%$ found it to be contracted. For neutral, there is a clear mode at ``neither contracted nor expanded'' (2). These statistics show that the users perceived the VAs to have clear preferences of the different intensities of the pose affective features when expressing different emotions.

We also perform a $k$-sample Anderson-Darling test~\ref{tab:anderson_darling_test} for each gait and each of the four broad affective features (and $k$ being the number of users) on the null hypothesis that all the user-provided values are from the same underlying distribution. We fail to reject the null hypothesis for all the four broad features in all the gaits, thus indicating strong statistical evidence that the observed intensities are consistent across different users.

\subsubsection{Other Feedback}
We asked the users to mark out of five how natural and smooth they felt the animations to be, with one indicating ``not natural at all'', three indicating ``satisfactory'', and five indicating ``very natural''. To establish a baseline, we asked the users to similarly marking the corresponding source motions as well. For our generated animations, 22\% of the users marked five, 43\% marked four, 24\% marked three, 7\% marked two, and 4\% marked one. Thus 89\% of the users marked at least three, \textit{i.e.}, found the naturalness in the generated gaits to be satisfactory. By contrast, for the source motions, 58\% of the users marked five, 35\% marked four, and 7\% marked three.

In the videos we sent out to the users, we used a moving camera so that the user was always looking straight at the virtual agent as it walked on different trajectories.
30\% of the users reported being distracted by this moving camera during the study. Therefore, we plan to use a fixed camera in our subsequent studies.

%% Future Work
\section{Conclusions, Limitations, and Future Work}
\label{sec:limitations}
We present a novel, learning-based method to synthesize and transition between emotive gaits. Our emotion model is based on a linear combination of four widely-used categorical emotions and we present a network architecture that uses affective features and movement features. Our algorithm can generate emotive gaits that follow a given trajectory at interactive rates and develops a transition scheme to switch between gaits with different emotions. We have shown the results on gaits collected from open-source datasets and discussed our procedure for developing VAs with these gaits in an AR environment. We have also reported our observations from a web-based user study to conclude that our generated gaits looked natural, as well as had the desired emotional expressiveness. Lastly, we release an augmented dataset of emotive gaits.

Our work has some limitations. Our approach can generate gaits of various emotions for one person at a time; it would be useful to generate gaits for a group of pedestrians in a crowd. Our formulation only considers a linear space of four emotions and we would like to extend our emotion representation to encompass more emotions in the arousal space and in the broader VAD space~\cite{vad}. The fidelity of our synthesized gaits is limited by the number of gaits and emotion labels available in the original database used by the network for training. To improve the performance and generate more natural-looking emotive gaits, we need larger datasets that account for individual, social, and cultural diversities. Moreover, our approach is only based on low-level affective and movement features, and it would be useful to model higher-level information corresponding to the environment and context. Furthermore, we would like to combine emotive gaits with other cues corresponding to facial expressions or gestures and use multiple modalities.

\acknowledgments{This work has been supported by ARO grant W911NF-19-1-0069.}

\clearpage

% The authors wish to thank A, B, and C. This work was supported in part by
% a grant from XYZ.}

%\bibliographystyle{abbrv}
\bibliographystyle{abbrv-doi}

\bibliography{main}

\begin{thebibliography}{10}

\bibitem{cmu_mocap}
Cmu graphics lab motion capture database.
\newblock {\em http://mocap.cs.cmu.edu/}, 2018.

\bibitem{neon}
{\em NEON: https://www.neon.life/}, 2020.

\bibitem{emotion_culture_diversity1}
J.~Altarriba, D.~M. Basnight, and T.~M. Canary.
\newblock Emotion representation and perception across cultures.
\newblock {\em Online readings in psychology and culture}, 4(1):1--17, 2003.

\bibitem{misleading_cues}
H.~Aviezer, Y.~Trope, and A.~Todorov.
\newblock Body cues, not facial expressions, discriminate between intense
  positive and negative emotions.
\newblock {\em Science}, 338(6111):1225--1229, 2012.

\bibitem{context_PS}
L.~F. Barrett, B.~Mesquita, and M.~Gendron.
\newblock Context in emotion perception.
\newblock {\em Current Directions in Psychological Science}, 20(5):286--290,
  2011. doi: {{%
10\hspace{.1pt}\discretionary{.}{%
}{.}\hspace{.4pt}1177\%2F0963721411422522}}


\bibitem{robotics}
A.~Bauer, K.~Klasing, G.~Lidoris, Q.~M{\"u}hlbauer, F.~Rohrm{\"u}ller,
  S.~Sosnowski, T.~Xu, K.~K{\"u}hnlenz, D.~Wollherr, and M.~Buss.
\newblock The autonomous city explorer: Towards natural human-robot interaction
  in urban environments.
\newblock {\em International Journal of Social Robotics}, 1(2):127--140, Apr
  2009. doi: {{%
10\hspace{.1pt}\discretionary{.}{%
}{.}\hspace{.4pt}1007\discretionary{/}{%
}{/}s12369\discretionary{%
}{-}{-}009\discretionary{%
}{-}{-}0011\discretionary{%
}{-}{-}9}}


\bibitem{curriculum_schedule}
S.~Bengio, O.~Vinyals, N.~Jaitly, and N.~Shazeer.
\newblock Scheduled sampling for sequence prediction with recurrent neural
  networks.
\newblock In C.~Cortes, N.~D. Lawrence, D.~D. Lee, M.~Sugiyama, and R.~Garnett,
  eds., {\em Advances in Neural Information Processing Systems 28}, pp.
  1171--1179. Curran Associates, Inc., 2015.

\bibitem{elu}
Y.~Bengio and Y.~LeCun, eds.
\newblock {\em 4th International Conference on Learning Representations, {ICLR}
  2016, San Juan, Puerto Rico, May 2-4, 2016, Conference Track Proceedings},
  2016.

\bibitem{step}
U.~Bhattacharya, T.~Mittal, R.~Chandra, T.~Randhavane, A.~Bera, and D.~Manocha.
\newblock Step: Spatial temporal graph convolutional networks for emotion
  perception from gaits.
\newblock In {\em Proceedings of the Thirty-Fourth AAAI Conference on
  Artificial Intelligence}, AAAI’20, p. 1342–1350. AAAI Press, 2020.

\bibitem{taew}
U.~Bhattacharya, C.~Roncal, T.~Mittal, R.~Chandra, A.~Bera, and D.~Manocha.
\newblock Take an emotion walk: Perceiving emotions from gaits using
  hierarchical attention pooling and affective mapping.
\newblock In {\em Proceedings of the European Conference on Computer Vision
  (ECCV)}, August 2020.

\bibitem{verbal_comm1}
A.~Chowanda, P.~Blanchfield, M.~Flintham, and M.~Valstar.
\newblock Computational models of emotion, personality, and social
  relationships for interactions in games: (extended abstract).
\newblock In {\em Proceedings of the 2016 International Conference on
  Autonomous Agents and Multiagent Systems}, AAMAS ’16, p. 1343–1344.
  International Foundation for Autonomous Agents and Multiagent Systems,
  Richland, SC, 2016.

\bibitem{face_and_posture}
C.~Clavel, J.~Plessier, J.-C. Martin, L.~Ach, and B.~Morel.
\newblock Combining facial and postural expressions of emotions in a virtual
  character.
\newblock In Z.~Ruttkay, M.~Kipp, A.~Nijholt, and H.~H. Vilhj{\'a}lmsson, eds.,
  {\em Intelligent Virtual Agents}, pp. 287--300. Springer Berlin Heidelberg,
  Berlin, Heidelberg, 2009.

\bibitem{crenn2016body}
A.~Crenn, R.~A. Khan, A.~Meyer, and S.~Bouakaz.
\newblock Body expression recognition from animated 3d skeleton.
\newblock In {\em IC3D}, pp. 1--7. IEEE, 2016.

\bibitem{dynamic_prediction}
Q.~Cui, H.~Sun, and F.~Yang.
\newblock Learning dynamic relationships for 3d human motion prediction.
\newblock In {\em Proceedings of the IEEE/CVF Conference on Computer Vision and
  Pattern Recognition (CVPR)}, June 2020.

\bibitem{posture_ex}
N.~Dael, M.~Mortillaro, and K.~R. Scherer.
\newblock Emotion expression in body action and posture.
\newblock {\em Emotion}, 12(5):1085, 2012. doi: {{%
10\hspace{.1pt}\discretionary{.}{%
}{.}\hspace{.4pt}1037\discretionary{/}{%
}{/}a0025737}}


\bibitem{style_cvae}
H.~Du, E.~Herrmann, J.~Sprenger, N.~Cheema, S.~Hosseini, K.~Fischer, and
  P.~Slusallek.
\newblock Stylistic locomotion modeling with conditional variational
  autoencoder.
\newblock In {\em Eurographics (Short Papers)}, pp. 9--12, 2019.

\bibitem{face_ex}
P.~Ekman and W.~V. Friesen.
\newblock {\em Facial action coding system: Investigator's guide}.
\newblock Consulting Psychologists Press, 1978.

\bibitem{expressive_face}
Y.~Ferstl and R.~McDonnell.
\newblock A perceptual study on the manipulation of facial features for trait
  portrayal in virtual agents.
\newblock In {\em Proceedings of the 18th International Conference on
  Intelligent Virtual Agents}, IVA ’18, p. 281–288. Association for
  Computing Machinery, New York, NY, USA, 2018. doi: {{%
10\hspace{.1pt}\discretionary{.}{%
}{.}\hspace{.4pt}1145\discretionary{/}{%
}{/}3267851\hspace{.1pt}\discretionary{.}{%
}{.}\hspace{.4pt}3267891}}


\bibitem{voice_ex}
R.~W. Frick.
\newblock Communicating emotion: The role of prosodic features.
\newblock {\em Psychological Bulletin}, 97(3):412, 1985. doi: {{%
10\hspace{.1pt}\discretionary{.}{%
}{.}\hspace{.4pt}1037\discretionary{/}{%
}{/}0033\discretionary{%
}{-}{-}2909\hspace{.1pt}\discretionary{.}{%
}{.}\hspace{.4pt}97\hspace{.1pt}\discretionary{.}{%
}{.}\hspace{.4pt}3\hspace{.1pt}\discretionary{.}{%
}{.}\hspace{.4pt}412}}


\bibitem{emotion_culture_diversity2}
M.~Gendron, D.~Roberson, J.~M. van~der Vyver, and L.~F. Barrett.
\newblock Perceptions of emotion from facial expressions are not culturally
  universal: evidence from a remote culture.
\newblock {\em Emotion}, 14(2):251, 2014. doi: {{%
10\hspace{.1pt}\discretionary{.}{%
}{.}\hspace{.4pt}1037\discretionary{/}{%
}{/}a0036052}}


\bibitem{rocketbox}
M.~Gonzalez-Franco, M.~Wojcik, E.~Ofek, A.~Steed, and D.~Garagan.
\newblock {\em Microsoft Rocketbox:
  https://github.com/microsoft/Microsoft-Rocketbox}, 2020.

\bibitem{effort_shape}
M.~M. Gross, E.~A. Crane, and B.~L. Fredrickson.
\newblock Effort-shape and kinematic assessment of bodily expression of emotion
  during gait.
\newblock {\em Human movement science}, 31(1):202--221, 2012.

\bibitem{elmd}
I.~Habibie, D.~Holden, J.~Schwarz, J.~Yearsley, and T.~Komura.
\newblock A recurrent variational autoencoder for human motion synthesis.
\newblock In {\em British Machine Vision Conference 2017, {BMVC} 2017, London,
  UK, September 4-7, 2017}, 2017.

\bibitem{pfnn}
D.~Holden, T.~Komura, and J.~Saito.
\newblock Phase-functioned neural networks for character control.
\newblock {\em ACM Transactions on Graphics (TOG)}, 36(4):42, 2017.

\bibitem{motion_synth}
D.~Holden, J.~Saito, and T.~Komura.
\newblock A deep learning framework for character motion synthesis and editing.
\newblock {\em ACM Transactions on Graphics (TOG)}, 35(4):138, 2016.

\bibitem{human3.6m}
C.~Ionescu, D.~Papava, V.~Olaru, and C.~Sminchisescu.
\newblock Human3.6m: Large scale datasets and predictive methods for 3d human
  sensing in natural environments.
\newblock {\em IEEE transactions on pattern analysis and machine intelligence},
  36(7):1325--1339, 2013.

\bibitem{bonding_in_conversations}
N.~Jaques, D.~J. McDuff, Y.~L. Kim, and R.~W. Picard.
\newblock Understanding and predicting bonding in conversations using thin
  slices of facial expressions and body language.
\newblock In D.~R. Traum, W.~R. Swartout, P.~Khooshabeh, S.~Kopp, S.~Scherer,
  and A.~Leuski, eds., {\em Intelligent Virtual Agents - 16th International
  Conference, {IVA} 2016, Los Angeles, CA, USA, September 20-23, 2016,
  Proceedings}, vol. 10011 of {\em Lecture Notes in Computer Science}, pp.
  64--74, 2016. doi: {{%
10\hspace{.1pt}\discretionary{.}{%
}{.}\hspace{.4pt}1007\discretionary{/}{%
}{/}978\discretionary{%
}{-}{-}3\discretionary{%
}{-}{-}319\discretionary{%
}{-}{-}47665\discretionary{%
}{-}{-}0}}


\bibitem{karg2013body}
M.~Karg, A.-A. Samadani, R.~Gorbet, K.~K{\"u}hnlenz, J.~Hoey, and D.~Kuli{\'c}.
\newblock Body movements for affective expression: A survey of automatic
  recognition and generation.
\newblock {\em IEEE Transactions on Affective Computing}, 4(4):341--359, 2013.

\bibitem{emotions_social}
D.~Keltner and J.~Haidt.
\newblock Social functions of emotions.
\newblock 2001.

\bibitem{unpaired_style_transfer}
A.~Kfir, Y.~Weng, D.~Lischinski, D.~Cohen-Or, and B.~Chen.
\newblock Unpaired motion style transfer from video to animation.
\newblock {\em ACM Trans. Graph.}, 39(4), July 2020. doi: {{%
10\hspace{.1pt}\discretionary{.}{%
}{.}\hspace{.4pt}1145\discretionary{/}{%
}{/}3386569\hspace{.1pt}\discretionary{.}{%
}{.}\hspace{.4pt}3392469}}


\bibitem{adam}
D.~P. Kingma and J.~Ba.
\newblock Adam: A method for stochastic optimization.
\newblock {\em arXiv preprint arXiv:1412.6980}, 2014.

\bibitem{context_based_ER}
R.~{Kosti}, J.~{Alvarez}, A.~{Recasens}, and A.~{Lapedriza}.
\newblock Context based emotion recognition using emotic dataset.
\newblock {\em IEEE Transactions on Pattern Analysis and Machine Intelligence},
  pp. 1--1, 2019. doi: {{%
10\hspace{.1pt}\discretionary{.}{%
}{.}\hspace{.4pt}1109\discretionary{/}{%
}{/}TPAMI\hspace{.1pt}\discretionary{.}{%
}{.}\hspace{.4pt}2019\hspace{.1pt}\discretionary{.}{%
}{.}\hspace{.4pt}2916866}}


\bibitem{motion_graphs}
L.~Kovar, M.~Gleicher, and F.~Pighin.
\newblock Motion graphs.
\newblock In {\em ACM SIGGRAPH 2008 Classes}, SIGGRAPH ’08. Association for
  Computing Machinery, New York, NY, USA, 2008. doi: {{%
10\hspace{.1pt}\discretionary{.}{%
}{.}\hspace{.4pt}1145\discretionary{/}{%
}{/}1401132\hspace{.1pt}\discretionary{.}{%
}{.}\hspace{.4pt}1401202}}


\bibitem{embodied_ambient_crowds}
M.~E. {Latoschik}, F.~{Kern}, J.~{Stauffert}, A.~{Bartl}, M.~{Botsch}, and
  J.~{Lugrin}.
\newblock Not alone here?! scalability and user experience of embodied ambient
  crowds in distributed social virtual reality.
\newblock {\em IEEE Transactions on Visualization and Computer Graphics},
  25(5):2134--2144, 2019.

\bibitem{avatar_realism}
M.~E. Latoschik, D.~Roth, D.~Gall, J.~Achenbach, T.~Waltemate, and M.~Botsch.
\newblock The effect of avatar realism in immersive social virtual realities.
\newblock In {\em Proceedings of the 23rd ACM Symposium on Virtual Reality
  Software and Technology}, VRST ’17. Association for Computing Machinery,
  New York, NY, USA, 2017. doi: {{%
10\hspace{.1pt}\discretionary{.}{%
}{.}\hspace{.4pt}1145\discretionary{/}{%
}{/}3139131\hspace{.1pt}\discretionary{.}{%
}{.}\hspace{.4pt}3139156}}


\bibitem{motion_patches}
K.~H. Lee, M.~G. Choi, and J.~Lee.
\newblock Motion patches: building blocks for virtual environments annotated
  with motion data.
\newblock {\em ACM Trans. Graph.}, 25(3):898--906, 2006.

\bibitem{muscleactuation}
S.~Lee, M.~Park, K.~Lee, and J.~Lee.
\newblock Scalable muscle-actuated human simulation and control.
\newblock {\em ACM Transactions on Graphics (TOG)}, 38(4):73, 2019.

\bibitem{seq2seq_dynamics}
C.~Li, Z.~Zhang, W.~S. Lee, and G.~H. Lee.
\newblock Convolutional sequence to sequence model for human dynamics.
\newblock In {\em Proceedings of the IEEE Conference on Computer Vision and
  Pattern Recognition (CVPR)}, June 2018.

\bibitem{generalized_combinations}
B.~Liebold and P.~Ohler.
\newblock Multimodal emotion expressions of virtual agents, mimic and vocal
  emotion expressions and their effects on emotion recognition.
\newblock In {\em Proceedings of the 2013 Humaine Association Conference on
  Affective Computing and Intelligent Interaction}, ACII ’13, p. 405–410.
  IEEE Computer Society, USA, 2013. doi: {{%
10\hspace{.1pt}\discretionary{.}{%
}{.}\hspace{.4pt}1109\discretionary{/}{%
}{/}ACII\hspace{.1pt}\discretionary{.}{%
}{.}\hspace{.4pt}2013\hspace{.1pt}\discretionary{.}{%
}{.}\hspace{.4pt}73}}


\bibitem{sl_survey}
Z.~C. Lipton, J.~Berkowitz, and C.~Elkan.
\newblock A critical review of recurrent neural networks for sequence learning.
\newblock {\em arXiv preprint arXiv:1506.00019}, 2015.

\bibitem{bml}
Y.~Ma, H.~M. Paterson, and F.~E. Pollick.
\newblock A motion capture library for the study of identity, gender, and
  emotion perception from biological motion.
\newblock {\em Behavior research methods}, 38(1):134--141, 2006.

\bibitem{few_shot_homogeneous}
I.~Mason, S.~Starke, H.~Zhang, H.~Bilen, and T.~Komura.
\newblock Few-shot learning of homogeneous human locomotion styles.
\newblock {\em Computer Graphics Forum}, 37(7):143--153, 2018. doi: {{%
10\hspace{.1pt}\discretionary{.}{%
}{.}\hspace{.4pt}1111\discretionary{/}{%
}{/}cgf\hspace{.1pt}\discretionary{.}{%
}{.}\hspace{.4pt}13555}}


\bibitem{body_posture_crowds}
J.~E. McHugh, R.~McDonnell, C.~O’Sullivan, and F.~N. Newell.
\newblock Perceiving emotion in crowds: the role of dynamic body postures on
  the perception of emotion in crowded scenes.
\newblock {\em Experimental brain research}, 204(3):361--372, 2010.

\bibitem{vad}
A.~Mehrabian and J.~A. Russell.
\newblock {\em An approach to environmental psychology.}
\newblock the MIT Press, 1974.

\bibitem{m3er}
T.~Mittal, U.~Bhattacharya, R.~Chandra, A.~Bera, and D.~Manocha.
\newblock M3er: Multiplicative multimodal emotion recognition using facial,
  textual, and speech cues.
\newblock In {\em Proceedings of the Thirty-Fourth AAAI Conference on
  Artificial Intelligence}, AAAI’20, pp. 1359--1367. AAAI Press, 2020.

\bibitem{emoticon}
T.~Mittal, P.~Guhan, U.~Bhattacharya, R.~Chandra, A.~Bera, and D.~Manocha.
\newblock Emoticon: Context-aware multimodal emotion recognition using frege's
  principle.
\newblock In {\em Proceedings of the IEEE/CVF Conference on Computer Vision and
  Pattern Recognition}, pp. 14234--14243, 2020.

\bibitem{montepare}
J.~M. Montepare, S.~B. Goldstein, and A.~Clausen.
\newblock The identification of emotions from gait information.
\newblock {\em Journal of Nonverbal Behavior}, 11(1):33--42, 1987.

\bibitem{small_group_interaction}
F.~Moustafa and A.~Steed.
\newblock A longitudinal study of small group interaction in social virtual
  reality.
\newblock In {\em Proceedings of the 24th ACM Symposium on Virtual Reality
  Software and Technology}, VRST ’18. Association for Computing Machinery,
  New York, NY, USA, 2018. doi: {{%
10\hspace{.1pt}\discretionary{.}{%
}{.}\hspace{.4pt}1145\discretionary{/}{%
}{/}3281505\hspace{.1pt}\discretionary{.}{%
}{.}\hspace{.4pt}3281527}}


\bibitem{ict}
S.~Narang, A.~Best, A.~Feng, S.-h. Kang, D.~Manocha, and A.~Shapiro.
\newblock Motion recognition of self and others on realistic 3d avatars.
\newblock {\em Computer Animation and Virtual Worlds}, 28(3-4):e1762, 2017.

\bibitem{proxemo}
V.~Narayanan, B.~M. Manoghar, V.~S. Dorbala, D.~Manocha, and A.~Bera.
\newblock Proxemo: Gait-based emotion learning and multi-view proxemic fusion
  for socially-aware robot navigation.
\newblock In {\em 2020 {IEEE/RSJ} International Conference on Intelligent
  Robots and Systems, {IROS} 2020}. {IEEE}, 2020.

\bibitem{depression}
E.~J. Nestler, M.~Barrot, R.~J. DiLeone, A.~J. Eisch, S.~J. Gold, and L.~M.
  Monteggia.
\newblock Neurobiology of depression.
\newblock {\em Neuron}, 34(1):13--25, 2002. doi: {{%
10\hspace{.1pt}\discretionary{.}{%
}{.}\hspace{.4pt}1016\discretionary{/}{%
}{/}S0896\discretionary{%
}{-}{-}6273\discretionary{%
}{(}{(}02\discretionary{)}{%
}{)}00653\discretionary{%
}{-}{-}0}}


\bibitem{games}
H.~Osking and J.~A. Doucette.
\newblock Enhancing emotional effectiveness of virtual-reality experiences with
  voice control interfaces.
\newblock In B.~{et al.}, ed., {\em Immersive Learning Research Network}, pp.
  199--209. Springer International Publishing, Cham, 2019.

\bibitem{icc}
S.~Park, H.~Ryu, S.~Lee, S.~Lee, and J.~Lee.
\newblock Learning predict-and-simulate policies from unorganized human motion
  data.
\newblock {\em ACM Trans. Graph.}, 38(6), 2019.

\bibitem{quaternet}
D.~Pavllo, D.~Grangier, and M.~Auli.
\newblock Quaternet: {A} quaternion-based recurrent model for human motion.
\newblock In {\em British Machine Vision Conference 2018, {BMVC} 2018}, p. 299,
  2018.

\bibitem{evaluate_recognition}
I.~Pelczer, F.~C. Contreras, and F.~G. Rodr{\'i}guez.
\newblock Expressions of emotions in virtual agents: Empirical evaluation.
\newblock {\em 2007 IEEE Symposium on Virtual Environments, Human-Computer
  Interfaces and Measurement Systems}, pp. 31--35, 2007.

\bibitem{deep_mimic}
X.~B. Peng, P.~Abbeel, S.~Levine, and M.~van~de Panne.
\newblock Deepmimic: Example-guided deep reinforcement learning of
  physics-based character skills.
\newblock {\em ACM Trans. Graph.}, 37(4), July 2018. doi: {{%
10\hspace{.1pt}\discretionary{.}{%
}{.}\hspace{.4pt}1145\discretionary{/}{%
}{/}3197517\hspace{.1pt}\discretionary{.}{%
}{.}\hspace{.4pt}3201311}}


\bibitem{deep_loco}
X.~B. Peng, G.~Berseth, K.~Yin, and M.~Van De~Panne.
\newblock Deeploco: Dynamic locomotion skills using hierarchical deep
  reinforcement learning.
\newblock {\em ACM Trans. Graph.}, 36(4), July 2017. doi: {{%
10\hspace{.1pt}\discretionary{.}{%
}{.}\hspace{.4pt}1145\discretionary{/}{%
}{/}3072959\hspace{.1pt}\discretionary{.}{%
}{.}\hspace{.4pt}3073602}}


\bibitem{neurobiology_of_perception}
M.~L. Phillips, W.~C. Drevets, S.~L. Rauch, and R.~Lane.
\newblock Neurobiology of emotion perception i: The neural basis of normal
  emotion perception.
\newblock {\em Biological psychiatry}, 54(5):504--514, 2003.

\bibitem{tanmay_emotions}
T.~Randhavane, A.~Bera, K.~Kapsaskis, U.~Bhattacharya, K.~Gray, and D.~Manocha.
\newblock Identifying emotions from walking using affective and deep features.
\newblock {\em arXiv preprint arXiv:1906.11884}, 2019.

\bibitem{fva}
T.~Randhavane, A.~Bera, K.~Kapsaskis, K.~Gray, and D.~Manocha.
\newblock Fva: Modeling perceived friendliness of virtual agents using movement
  characteristics.
\newblock {\em IEEE transactions on visualization and computer graphics},
  25(11):3135--3145, 2019.

\bibitem{eva}
T.~Randhavane, A.~Bera, K.~Kapsaskis, R.~Sheth, K.~Gray, and D.~Manocha.
\newblock Eva: Generating emotional behavior of virtual agents using expressive
  features of gait and gaze.
\newblock In {\em ACM Symposium on Applied Perception 2019}, p.~6. ACM, 2019.

\bibitem{dominance}
T.~Randhavane, A.~Bera, E.~Kubin, K.~Gray, and D.~Manocha.
\newblock Modeling data-driven dominance traits for virtual characters using
  gait analysis.
\newblock {\em CoRR}, abs/1901.02037, 2019.

\bibitem{liarswalk}
T.~Randhavane, U.~Bhattacharya, K.~Kapsaskis, K.~Gray, A.~Bera, and D.~Manocha.
\newblock The liar's walk: Detecting deception with gait and gesture.
\newblock {\em arXiv preprint arXiv:1912.06874}, 2019.

\bibitem{anthropomorphism}
L.~D. Riek, T.-C. Rabinowitch, B.~Chakrabarti, and P.~Robinson.
\newblock How anthropomorphism affects empathy toward robots.
\newblock In {\em Proceedings of the 4th ACM/IEEE International Conference on
  Human Robot Interaction}, HRI ’09, p. 245–246. Association for Computing
  Machinery, New York, NY, USA, 2009. doi: {{%
10\hspace{.1pt}\discretionary{.}{%
}{.}\hspace{.4pt}1145\discretionary{/}{%
}{/}1514095\hspace{.1pt}\discretionary{.}{%
}{.}\hspace{.4pt}1514158}}


\bibitem{therapy}
J.~J. Rivas, F.~Orihuela-Espina, L.~E. Sucar, L.~Palafox,
  J.~Hern\'{a}ndez-Franco, and N.~Bianchi-Berthouze.
\newblock Detecting affective states in virtual rehabilitation.
\newblock In {\em Proceedings of the 9th International Conference on Pervasive
  Computing Technologies for Healthcare}, PervasiveHealth ’15, p. 287–292.
  ICST (Institute for Computer Sciences, Social-Informatics and
  Telecommunications Engineering), Brussels, BEL, 2015.

\bibitem{critical_gait_features}
C.~L. Roether, L.~Omlor, A.~Christensen, and M.~A. Giese.
\newblock Critical features for the perception of emotion from gait.
\newblock {\em Journal of vision}, 9(6):15--15, 2009.

\bibitem{biped}
N.~Rokbani, B.~A. Cherif, and A.~M. Alimi.
\newblock Toward intelligent biped-humanoids gaits generation.
\newblock {\em Humanoid Robots}, pp. 259--271, 2009.

\bibitem{brain_injury}
H.~Rosenberg, S.~McDonald, J.~Rosenberg, and R.~F. Westbrook.
\newblock Measuring emotion perception following traumatic brain injury: The
  complex audio visual emotion assessment task (caveat).
\newblock {\em Neuropsychological Rehabilitation}, 29(2):232--250, 2019.
\newblock PMID: 28030989. doi: {{%
10\hspace{.1pt}\discretionary{.}{%
}{.}\hspace{.4pt}1080\discretionary{/}{%
}{/}09602011\hspace{.1pt}\discretionary{.}{%
}{.}\hspace{.4pt}2016\hspace{.1pt}\discretionary{.}{%
}{.}\hspace{.4pt}1273118}}


\bibitem{anderson_darling_ks}
F.~W. Scholz and M.~A. Stephens.
\newblock K-sample anderson–darling tests.
\newblock {\em Journal of the American Statistical Association},
  82(399):918--924, 1987. doi: {{%
10\hspace{.1pt}\discretionary{.}{%
}{.}\hspace{.4pt}1080\discretionary{/}{%
}{/}01621459\hspace{.1pt}\discretionary{.}{%
}{.}\hspace{.4pt}1987\hspace{.1pt}\discretionary{.}{%
}{.}\hspace{.4pt}10478517}}


\bibitem{verbal_comm2}
S.~S. Sohn, X.~Zhang, F.~Geraci, and M.~Kapadia.
\newblock An emotionally aware embodied conversational agent.
\newblock In {\em Proceedings of the 17th International Conference on
  Autonomous Agents and MultiAgent Systems}, AAMAS ’18, p. 2250–2252.
  International Foundation for Autonomous Agents and Multiagent Systems,
  Richland, SC, 2018.

\bibitem{vr}
B.~Stangl, D.~C. Ukpabi, and S.~Park.
\newblock Augmented reality applications: The impact of usability and emotional
  perceptions on tourists' app experiences.
\newblock In J.~Neidhardt and W.~W{\"o}rndl, eds., {\em Information and
  Communication Technologies in Tourism 2020}, pp. 181--191. Springer
  International Publishing, Cham, 2020.

\bibitem{nsm}
S.~Starke, H.~Zhang, T.~Komura, and J.~Saito.
\newblock Neural state machine for character-scene interactions.
\newblock {\em ACM Transactions on Graphics (TOG)}, 38(6):209, 2019.

\bibitem{kernel_based2}
J.~M. Wang, D.~J. Fleet, and A.~Hertzmann.
\newblock Gaussian process dynamical models for human motion.
\newblock {\em IEEE transactions on pattern analysis and machine intelligence},
  30(2):283--298, 2007.

\bibitem{style_transfer_heterogeneous}
S.~Xia, C.~Wang, J.~Chai, and J.~Hodgins.
\newblock Realtime style transfer for unlabeled heterogeneous human motion.
\newblock {\em ACM Trans. Graph.}, 34(4), July 2015. doi: {{%
10\hspace{.1pt}\discretionary{.}{%
}{.}\hspace{.4pt}1145\discretionary{/}{%
}{/}2766999}}


\bibitem{spectral_style_transfer}
M.~E. Yumer and N.~J. Mitra.
\newblock Spectral style transfer for human motion between independent actions.
\newblock {\em ACM Trans. Graph.}, 35(4), July 2016. doi: {{%
10\hspace{.1pt}\discretionary{.}{%
}{.}\hspace{.4pt}1145\discretionary{/}{%
}{/}2897824\hspace{.1pt}\discretionary{.}{%
}{.}\hspace{.4pt}2925955}}


\end{thebibliography}
\end{document}